\documentclass{nic-series}
\usepackage{graphicx,epsfig,cite,overcite}
\usepackage{bbm,amssymb}
\def\eka{[}
\def\ekz{]}

\begin{document} 

\title{The LDA+DMFT Approach to Materials with Strong Electronic
Correlations}

\author{K. Held \inst{1},  I.~A.~Nekrasov \inst{2}, G.~Keller \inst{3},  V.~Eyert \inst{4}, N.~Bl{\"{u}}mer \inst{5},
A.~K.~McMahan \inst{6}, R.~T.~Scalettar \inst{7}, T.~Pruschke  \inst{3}, V.~I.~Anisimov  \inst{2}, and D.~Vollhardt \inst{3}}

\institute{ Physics
Department, Princeton University, Princeton, NJ 08544, USA\\
         \and
        Institute of Metal Physics, Russian Academy of Sciences-Ural Division,\\
Yekaterinburg GSP-170, Russia\\
         \and
        Theoretical Physics III, Center
        for Electronic Correlations and Magnetism,\\
        Institute for Physics,  University of Augsburg, D-86135 Augsburg, Germany\\
         \and
        Institute for Physics, Theoretical Physics II,   University of Augsburg, \\  D-86135 Augsburg, Germany\\
        \and
        Institute for Physics,
Johannes Gutenberg University, 
D-55099 Mainz, Germany\\
        \and
        Lawrence Livermore National Laboratory, University of California,
Livermore, CA 94550, USA\\
        \and
        Physics Department, University of California, Davis, CA 95616, USA
          }

\maketitle

\begin{abstracts}
LDA+DMFT is a novel computational technique for {\em ab initio}
investigations of real materials with strongly correlated electrons,
such as transition metals and their oxides. It combines the strength of
conventional band structure theory in the local density approximation
(LDA) with a modern many-body approach, the dynamical mean-field theory
(DMFT). In the last few years LDA+DMFT has proved to be a powerful tool
for the realistic modeling of strongly correlated electronic systems.
In this paper the basic ideas and the
set-up of the LDA+DMFT(X) approach, where X is the method used to
solve the DMFT equations, are discussed. Results obtained with
X=QMC (quantum Monte Carlo) and X=NCA (non-crossing approximation)
are presented and compared. By means of the model system
La$_{1-x}$Sr$_{x}$TiO$_{3}$ we show that the method X matters qualitatively
and quantitatively. Furthermore, we discuss recent results on the
Mott-Hubbard metal-insulator transition in the transition metal oxide 
V$_2$O$_3$ and
%% BEGIN LaMnO3
%%, on orbital ordering and magnetism in LaMnO$_3$,
%% END LaMnO3
the $\alpha$-$\gamma$ transition in the
4f-electron system Ce.
%% BEGIN XRuO4
%% and the evolution
%%of the electronic structure at intermediate concentrations ${\mathbf
%%(2>x>0.5)}$
%%in the isoelectric alloy series Ca$_{2-x}$Sr$_x$RuO$_4$.
%% END XRuO4
\end{abstracts}
\vspace{-.4cm}

%\tableofcontents
\begin{center}
Table of contents\\[0.15cm]
{\small
\begin{tabular}{l@{\ \ }p{0.85\textwidth}@{\ \ }r}
\ref{intro}.&Introduction\dotfill&  \pageref{intro}\\
\ref{lda_dmft}.& The LDA+DMFT approach\dotfill &  \pageref{lda_dmft}\\
&\ref{dft}.\ \ Density functional theory\dotfill & \pageref{dft}\\
&\ref{lda}.\ \ Local density approximation\dotfill & \pageref{lda}\\
&\ref{ldaCoulomb}.\ \ Supplementing LDA with local Coulomb correlations\dotfill&  \pageref{ldaCoulomb}\\
&\ref{dmft}.\ \ Dynamical mean-field theory\dotfill & \pageref{dmft}\\
&\ref{QMC}.\ \ QMC method to solve DMFT\dotfill & \pageref{QMC}\\
&\ref{NCA}.\ \ NCA method to solve DMFT\dotfill & \pageref{NCA}\\
&\ref{SimpTMO}.\ \ Simplifications for transition metal oxides with well separated
  $e_{g}$- and $t_{2g}$-bands\dotfill & \pageref{SimpTMO}\\
&\ref{selfconsist}.\ \ Self-consistent LDA+DMFT\dotfill & \pageref{selfconsist}\\
\ref{LaTiO3}.&Comparison of different methods to solve DMFT\dotfill &
\pageref{LaTiO3}\\ 
\ref{v2o3}.&Mott-Hubbard metal-insulator transition in V$_2$O$_3$\dotfill 
&\pageref{v2o3}\\
\ref{Ce}.&The Cerium volume collapse\dotfill 
& \pageref{Ce}\\
%% BEGIN LaMnO3
%\ref{LaMnO3}.&Beyond paramagnetic phases: Orbital ordering and magnetism in
%LaMnO$_3$\dotfill &\pageref{LaMnO3}\\
%% END LaMnO3
%% BEGIN XRuO4
%\ref{ruthenates}.&Evolution of electronic structure in
%Ca$_{2-x}$Sr$_x$RuO$_4$ ${\mathbf (2>x>0.5)}$
%\dotfill
%&\pageref{ruthenates}\\
%% END XRuO4
\ref{conclusion}.&Conclusion and Outlook\dotfill & \pageref{conclusion}\\[.2cm]
\end{tabular}}
\end{center}

\hrulefill

\noindent To be published in the Proceedings of the Winter School on "Quantum
Simulations of Complex Many-Body Systems: From Theory to Algorithms",
February 25 - March 1, 2002, Rolduc/Kerkrade (NL), organized by the John
von Neumann Institute of Computing at the Forschungszentrum J\"ulich.

\section{Introduction}\label{intro}
The calculation of physical properties of electronic systems by
controlled approximations is one of the most important challenges
of modern theoretical solid state physics. In particular, the
physics of transition metal oxides -- a singularly important group
of materials both from the point of view of fundamental research
and technological applications -- may only be understood by
explicit consideration of the strong effective interaction between
the conduction electrons in these systems. The investigation of
electronic many-particle systems is made especially complicated by
quantum statistics, and by the fact that the investigation of many
phenomena require the application of non-perturbative theoretical
techniques.

From a microscopic point of view theoretical solid state physics
is concerned with the investigation of interacting many-particle
systems involving electrons and ions. However, it is an
established fact that many electronic properties of matter are
well described by the purely electronic Hamiltonian

\begin{eqnarray}
\hat{H} &=&\sum_{\sigma }\int \!d^{3}r\;\hat{\Psi}^{+}({\bf
r},\sigma )\left[
{-\frac{\hbar ^{2}}{2m_{e}}\Delta +V_{{\rm ion}}({\bf r})}\right] \hat{\Psi}(%
{\bf r},\sigma )  \nonumber \\
&&+\frac{1}{2}\sum_{\sigma \sigma ^{\prime }}\int \!d^{3}r \, d^{3}r^{\prime }\;
\hat{\Psi}^{+}({\bf r},\sigma )\hat{\Psi}^{+}({\bf r^{\prime
}},\sigma
^{\prime })\;{V_{{\rm ee}}({\bf r}\!-\!{\bf r^{\prime }})}\;\hat{\Psi}({\bf %
r^{\prime }},\sigma ^{\prime })\hat{\Psi}({\bf r},\sigma ),
\label{abinitioHam}
\end{eqnarray}%
where the crystal lattice enters only through an ionic potential.
The applicability of this approach may be justified by the
validity of the Born and Oppenheimer approximation\cite{Born27a}.
Here, $ \hat{\Psi}^{+}({\bf r},\sigma )$ and $\hat{\Psi}({\bf
r},\sigma )$ are field operators that create and annihilate an
electron at position ${\bf r}$ with spin $\sigma $, $\Delta $ is
the Laplace operator, $m_{e}$ the electron mass, $e$ the electron
charge, and
\begin{eqnarray}
V_{{\rm ion}} ({\bf r})=-e^2\sum_i \frac{Z_i}{|{\bf r}-{\bf R_i}|}
&\;\; {\rm and}\;\; & V_{\rm ee}({\bf r}\!-\!{\bf r'})=\frac{e^2}{2}
\sum_{{\bf r} \neq {\bf r'}} \frac{1}{|{\bf r}-{\bf r'}|}
\end{eqnarray}
denote the one-particle potential due to all ions $i$
with charge $eZ_{i}$ at given positions ${\bf R_{i}}$, and the
electron-electron interaction, respectively.

While the {\em ab initio} Hamiltonian (\ref{abinitioHam}) is easy
to write down it is impossible to solve exactly if more
than a few electrons are involved.
Numerical methods like Green's Function Monte Carlo and related 
approaches have been used successfully for relatively
modest numbers of electrons.  Even so, however, the focus of the 
work has been on jellium and on light atoms and molecules
like H, H$_2$, $^3$He, $^4$He, see, e.g., 
the articles by Anderson, Bernu, Ceperley {\it et al.} 
in the present Proceedings of the {\em NIC Winterschool 2002}.
Because of this, one generally either needs to make substantial
approximations to deal with the Hamiltonian (\ref{abinitioHam}),
or replace it by a greatly simplified model Hamiltonian. At
present these two different strategies for the investigation of the
electronic properties of solids are applied by two largely
separate groups: the density functional theory (DFT) and the
many-body community. It is known for a long time already that DFT,
together with its local density approximation (LDA), is a highly
successful technique for the calculation of the electronic
structure of many real materials\cite{JonesGunn}. However, for
strongly correlated materials, i.e., $d$- and $f$-electron systems
which have a Coulomb interaction comparable to the band-width,
DFT/LDA is seriously restricted in its accuracy and reliability.
Here, the model Hamiltonian approach is more general and powerful
since there exist systematic theoretical techniques to investigate
the many-electron problem with increasing accuracy. These
many-body techniques allow one to describe qualitative tendencies and
understand the basic mechanism of various physical phenomena. At
the same time the model Hamiltonian approach is seriously
restricted in its ability to make quantitative predictions since
the input parameters are not accurately known and hence need to be
adjusted. One of the most successful techniques in this respect is
the dynamical mean-field theory (DMFT) -- a non-perturbative
approach to strongly
correlated electron systems which was developed during the past decade\cite%
{MetzVoll89,MuHa89,Brandt,Vaclav,Georges92,Jarrell92,vollha93,pruschke,georges96}. The LDA+DMFT approach,
which was first formulated by Anisimov {\it et al.}\cite{poter97,Held01},
combines the strength of DFT/LDA to describe the weakly correlated
part of the {\em ab initio} Hamiltonian (\ref{abinitioHam}), i.e.,
electrons in $s$- and $p$-orbitals as well as the long-range
interaction of the $d$- and $f$-electrons, with the power of DMFT
to describe the strong correlations induced by the local Coulomb
interaction of the $d$- or $f$-electrons.

Starting from the {\em ab initio} Hamiltonian (\ref{abinitioHam}),
the LDA+DMFT approach is presented in Section \ref{lda_dmft}, including the DFT in 
Section \ref{dft}, the LDA in Section \ref{lda}, the construction of a model
Hamiltonian in Section \ref{ldaCoulomb}, and the DMFT in Section \ref{dmft}.
As methods used to solve the DMFT we discuss the  quantum Monte Carlo (QMC)
algorithm in Section \ref{QMC} and
the non-crossing approximation (NCA) in Section \ref{NCA}. A  simplified 
treatment for transition metal oxides is introduced in Section \ref{SimpTMO},
and the scheme of a self-consistent LDA+DMFT in Section \ref{selfconsist}.
As a particular example, the LDA+DMFT calculation for La$_{1-x}$Sr$_{x}$TiO$_{3}$
is discussed in Section \ref{LaTiO3}, emphasizing that the method X 
to solve the DMFT matters on a quantitative level. 
Our calculations for the Mott-Hubbard metal-insulator transition in V$_2$O$_3$
are presented in Section \ref{v2o3}, in comparison to the experiment.
Section \ref{Ce} reviews our recent calculations of the Ce $\alpha$-$\gamma$
transition, in the perspective of the models referred to as Kondo
volume collapse and Mott transition scenario.
%% BEGIN XRuO4
%% BEGIN LaMnO3
%Besides paramagnetic properties the DMFT also allows one to study broken-symmetry
%phases. As a simple yet important example we discuss 
%Orbital ordering and
%magnetism of LaMnO$_3$ is studied in Section \ref{LaMnO3},
%as an example of a calculation within a symmetry broken phase.
%% END LaMnO3
%In Section \ref{ruthenates} we present results for the qualitative 
%understanding of
%the evolution of the electronic structure in the isoelectric alloys series
%Ca$_{2-x}$Sr$_x$RuO$_4$ for intermediate doping ${\mathbf (2>x>0.5)}$.
%% END XRuO4
A discussion of the LDA+DMFT
approach and its
future prospects  in  Section~\ref{conclusion} closes the presentation.

\section{The LDA+DMFT approach}
\label{lda_dmft}

\subsection{Density functional theory}
\label{dft}

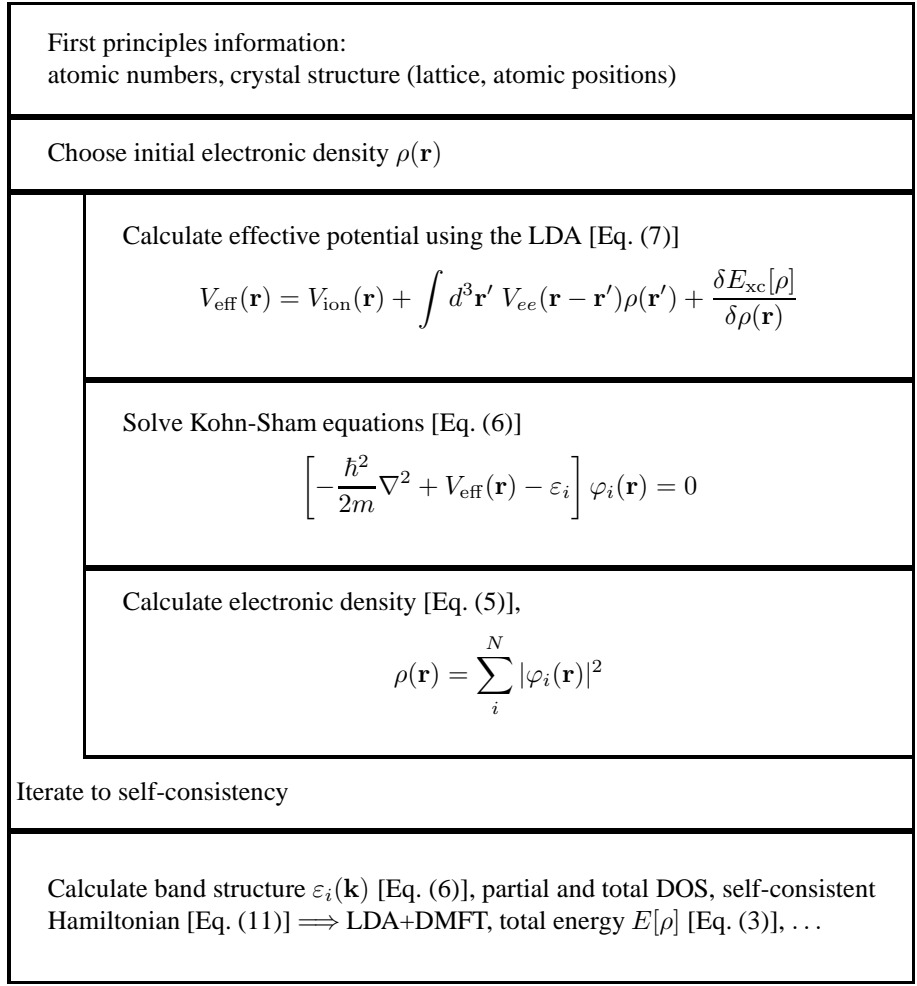
\begin{figure}[tb]
%
%     Flow diagram of the intraatomic calculations of the 
%     standard ASW method  by  Volker Eyert 
%
\begin{center}
\unitlength1mm
\begin{picture}(120,130)
\thicklines
\put(0.00,115.00){\framebox(120.00,15.00)[cc]
{\parbox{110mm}{ First principles information: \\ 
                 atomic numbers, 
                 crystal structure (lattice, atomic positions) }}}

\put(0.00,105.00){\framebox(120.00,10.00)[cc]
{\parbox{110mm}{ Choose initial electronic density 
                 $ \rho ( {\bf r} ) $}}}

\put(0.00,20.00){\framebox(120.00,85.00)[cc]{}}

\put(10.00,80.00){\framebox(110.00,25.00)[cc]
{\parbox{100mm}{ Calculate effective potential using the LDA 
                 [Eq.\ (\ref{VDFT})] 
                 \[
                 V_{\rm eff} ({\bf r})
                   =   V_{\rm ion} ({\bf r}) 
                       + \int d^3 {\bf r}' \;
                            V_{ee} ({\bf r} - {\bf r}') \rho ({\bf r}')
                       + \frac{ \delta E_{\rm xc} [ \rho ] }
                              { \delta \rho ({\bf r}) } 
                 \]
                 }}}

\put(10.00, 55.00){\framebox(110.00,25.00)[cc]
{\parbox{100mm}{ Solve Kohn-Sham equations [Eq.\ (\ref{KohnSham})]
                 \[
                 \left[ - \frac{ \hbar^2 }{ 2 m } \nabla^2 + V_{\rm eff} ({\bf r})
                        - \varepsilon_i
                 \right] \varphi_i ({\bf r})
                    = 0
                 \]
               }}}

\put(10.00, 30.00){\framebox(110.00,25.00)[cc]
{\parbox{100mm}{ Calculate electronic density [Eq.\ (\ref{rhophi})],  
                 \[
                 \rho ({\bf r})   =   \sum_i^N | \varphi_i ({\bf r}) |^2 
                 \]
               }}}

\put(0.00,24.00){
\parbox{80mm}{ Iterate to self-consistency }}

\put(0.00, 0.00){\framebox(120.00,20.00)[cc]
{\parbox{110mm}{ Calculate band structure $ \varepsilon_i ({\bf k}) $ 
                 [Eq.\ (\ref{KohnSham})],  
                 partial and total DOS, 
                 self-consistent Hamiltonian [Eq.\ (\ref{HLDA})] 
                 $ \Longrightarrow $ LDA+DMFT,  
                 total energy $ E [ \rho ] $ [Eq.\ (\ref{ELDA})],  
                 $ \ldots $}}}

\end{picture}
\end{center}
\caption{Flow diagram of the DFT/LDA calculations.} 
\label{fig:asws41} 
\end{figure}

The fundamental theorem of DFT by Hohenberg and
Kohn\cite{Hohenberg64a} 
%%% begin VHWGE
(see, e.g., the review by Jones and Gunnarsson\cite{JonesGunn})  
%%% end VHWGE
states that the ground state energy is a
functional of the electron density which assumes its minimum at
the ground state electron density.
 Following
Levy,\cite{Levy} this theorem is easily proved and the functional
even constructed by taking the minimum (infimum) of the energy
expectation value w.r.t. all (many-body) wave functions $\varphi
({\bf r_{1}}\sigma _{1},...\,{\bf r_{N}}\sigma _{N})$ at a given
electron number $N$ which yield the electron density $\rho ({\bf
r})$:
\begin{equation}
E[\rho ]=\inf \Big\{\langle \varphi |\hat{H}|\varphi \rangle \;\;\Big|%
\;\;\langle \varphi |\sum_{i=1}^{N}\delta ({\bf r}-{\bf
r_{i}})|\varphi \rangle =\rho ({\bf r})\Big\}.
\label{ELDA}
\end{equation}%
However, this construction is of no practical value since it
actually requires the evaluation of the Hamiltonian
(\ref{abinitioHam}). Only certain
contributions like the Hartree energy $E_{{\rm Hartree}}[\rho ]=\frac{1}{2}%
\int d^{3}r^{\prime }\,d^{3}r\;V_{{\rm ee}}({\bf r}\!-\!{\bf r^{\prime }}%
)\;\rho ({\bf r^{\prime })\rho (r)}$ and the energy of the ionic potential $%
E_{{\rm ion}}[\rho ]=\int d^{3}r\;V_{{\rm ion}}({\bf r})\;\rho
({\bf r)}$ can be expressed directly in terms of the electron
density. This leads to
\begin{equation}
E[\rho ]=E_{{\rm kin}}[\rho ]+E_{{\rm ion}}[\rho ]+E_{{\rm
Hartree}}[\rho ]+E_{{\rm xc}}[\rho ],  \label{Erho}
\end{equation}%
where $E_{{\rm kin}}[\rho ]$ denotes the kinetic energy, and $E_{{\rm xc}%
}[\rho ]$ is the unknown exchange and correlation term which
contains the energy of the electron-electron interaction except
for the Hartree term. Hence all the difficulties of the many-body
problem have been transferred into $E_{{\rm xc}}[\rho ]$. While the
kinetic energy $E_{{\rm kin}}$ cannot be expressed explicitly in
terms of the electron density one can employ a trick to determine
it. Instead of minimizing $E[\rho ]$ with respect to $\rho $ one minimizes
it w.r.t. a set of one-particle wave functions $\varphi _{i}$
related to $\rho $ via
\begin{equation}
\rho ({\bf r})=\sum_{i=1}^{N}|\varphi _{i}({\bf r})|^{2}.
\label{rhophi}
\end{equation}%
To guarantee the normalization of $\varphi_{i}$, the Lagrange parameters $%
\varepsilon _{i}$ are introduced such that the variation $\delta
\{E[\rho ]+\varepsilon _{i}[1-\int d^{3}r|\varphi _{i}({\bf
r})|^{2}]\}/\delta \varphi _{i}({\bf r})=0$ yields the
Kohn-Sham\cite{Kohn65} equations:
\begin{equation}
\left[ -{\frac{\hbar ^{2}}{2m_{e}}\Delta +V_{{\rm ion}}({\bf r})}+
\int d^{3}{r^{\prime }}\,
{V_{{\rm ee}}({\bf r}\!-\!{\bf r^{\prime }})} {\rho ({\bf r^{\prime }})}
+{{\frac{\delta {E_{{\rm xc}}[\rho ]}}{\delta \rho ({\bf r)}}}}%
\right] \varphi _{i}({\bf r})=\varepsilon _{i}\;\varphi _{i}({\bf
r}). \label{KohnSham}
\end{equation}%
These equations have the same form as a one-particle
Schr\"{o}dinger equation which, {\em a posteriori}, justifies to
calculate the kinetic energy by means of the one-particle
wave-function ansatz. The kinetic energy of 
a {\em one-particle} ansatz which has
the ground state density is, then, given by 
$E_{{\rm kin}}[\rho _{{\rm min%
}}]=-\sum_{i=1}^{N}\langle \varphi _{i}|{\hbar ^{2}\Delta }/{(2m_{e})}%
|\varphi _{i}\rangle $ if the $\varphi _{i}$ are the
self-consistent (spin-degenerate) solutions of Eqs.
(\ref{KohnSham}) and (\ref{rhophi}) with lowest ``energy''
$\epsilon_{i}$. Note, however, that the one-particle potential of
Eq. (\ref{KohnSham}), i.e.,
\begin{equation}
V_{\rm eff}({\bf r})=V_{{\rm ion}}({\bf r})+\int d^{3}{r^{\prime }}
{V_{{\rm ee}}({\bf r}\!-\!{\bf r^{\prime }})}{\rho ({\bf r^{\prime }})}
 +{{\frac{\delta {E_{{\rm xc}%
}[\rho ]}}{\delta \rho ({\bf r)}}}}, \label{VDFT}
\end{equation}%
is only an auxiliary potential which artificially arises in the
approach to minimize $E[\rho ]$. Thus, the wave functions $\varphi
_{i}$ and the Lagrange parameters $\varepsilon _{i}$ have no
physical meaning at this point.
Altogether, these equations allow for the DFT/LDA calculation,
see the flow diagram Fig.~\ref{fig:asws41}.

\subsection{Local density approximation}
\label{lda}
 So far no approximations have been employed since the
difficulty
of the many-body problem was only transferred to the unknown functional $%
E_{\rm xc}[\rho ]$. For this term the local density approximation
(LDA) which approximates the functional $E_{\rm xc}[\rho ]$ by a
function that depends on the local density only, i.e.,
\begin{equation}
E_{\rm xc}[\rho ]\;{\rightarrow }\;\int d^{3}r\;E_{\rm xc}^{{\rm LDA}}(\rho ({\bf r}%
)),
\end{equation}%
was found to be unexpectedly successful. Here, $E_{\rm xc}^{{\rm LDA}}(\rho (%
{\bf r}))$ is usually calculated from the  perturbative solution\cite{jellium}
 or the numerical
simulation\cite{jellium2} of the jellium problem which is defined 
by $V_{{\rm ion}}({\bf r})={\rm const}$.

In principle DFT/LDA only allows one to calculate static
properties like the ground state energy or its derivatives.
However, one of the major applications of LDA is the calculation
of band structures. To this end, the Lagrange parameters
$\varepsilon _{i}$ {\em are interpreted} as the physical
(one-particle) energies of the system under consideration. Since
the true ground-state is not a simple one-particle wave-function,
this is an approximation beyond DFT. Actually, this
approximation corresponds to the replacement of the Hamiltonian (\ref%
{abinitioHam}) by
\begin{eqnarray}
\hat{H}_{{\rm LDA}} &=&\sum_{\sigma }\int \!d^{3}r\;\hat{\Psi}^{+}({\bf r}%
,\sigma )\left[ -\frac{\hbar ^{2}}{2m_{e}}\Delta +V_{{\rm ion}}({\bf r}%
)+\int d^3{r^{\prime }}\,{\rho ({\bf r^{\prime }})}{V_{{\rm ee}}({\bf r}\!-\!%
{\bf r^{\prime }})}\right.   \nonumber \\ &&\left. \phantom
{\sum_{\sigma \sigma'} \int\! d^3r\; \hat{\Psi}^+({\bf r},\sigma)
\big[ \;}+{{\frac{\delta {E_{\rm xc}^{{\rm LDA}}}[\rho ]}{\delta \rho
({\bf r)}}}}\right] \hat{\Psi}({\bf r},\sigma ).  \label{HLDA0}
\end{eqnarray}%
For practical calculations one needs to expand the field operators
w.r.t. a basis $\Phi _{ilm}$, e.g., a linearized muffin-tin
orbital (LMTO)\cite{LMTO} basis ($i$ denotes lattice sites;
$l$ and $m$ are orbital indices). In this basis,
\begin{eqnarray}
\hat{\Psi}^+({\bf r},\sigma) &=& \sum_{i l m} \hat{c}_{i l
m}^{\sigma \dagger} \Phi_{i l m}({\bf r})^{\phantom{+}}
\end{eqnarray}
such that the Hamiltonian (\ref{HLDA0}) reads
\begin{eqnarray}
 \hat{H}_{\rm LDA}  &=& \sum_{ilm,{\rm }jl^{\prime }m^{\prime },\sigma }(\delta
_{ilm,jl^{\prime }m^{\prime }} \;
{\varepsilon_{ilm}}^{\phantom{\sigma}}{\hat{n}}_{ilm}^{\sigma
}+ {t_{ilm,jl^{\prime }m^{\prime }}}\;{\hat{c}}_{ilm}^{\sigma \dagger }{\hat{c}}%
_{jl^{\prime }m^{\prime }}^{\sigma }). \label{HLDA}
\end{eqnarray}
Here, ${\hat{n}}_{ilm}^{\sigma }=\hat{c}_{ilm}^{\sigma \dagger }{\hat{c}}%
_{ilm}^{\sigma }$,
\begin{equation}
t_{ilm,jl^{\prime }m^{\prime }}=\Big\langle\Phi
_{ilm}\Big|-\frac{\hbar
^{2}\Delta }{2m_{e}}+V_{{\rm ion}}({\bf r})+\int d^3{r^{\prime }}{\rho ({\bf %
r^{\prime }})}{V_{{\rm ee}}({\bf r}\!-\!{\bf r^{\prime }})}+{{\frac{\delta {%
E_{\rm xc}^{{\rm LDA}}}[\rho ]}{\delta \rho ({\bf r)}}}}\Big|\Phi
_{jl^{\prime }m^{\prime }}\Big\rangle
\end{equation}%
for $ilm\neq jl^{\prime }m^{\prime }$ and zero otherwise; $\varepsilon _{ilm}$
denotes the corresponding diagonal part.

As for static properties, the LDA approach
based on the self-consistent solution of Hamiltonian (\ref{HLDA}) 
together with the calculation
of the electronic density Eq.\ (\ref{rhophi}) [see the flow diagram
Fig.~\ref{fig:asws41}] has also been
highly successful for band structure calculations -- but only for weakly correlated materials\cite{JonesGunn}. It is not reliable when applied to correlated
materials and can even be completely wrong because it
  treats electronic {\em correlations} only very rudimentarily.
For example, it
predicts the antiferromagnetic insulator La$_{2}$CuO$_{4}$ to be a
non-magnetic metal\cite{LDAfailure} and also completely fails to account for the
high effective masses observed in $4f$-based heavy fermion compounds.

\subsection{Supplementing LDA with local Coulomb correlations}
\label{ldaCoulomb}

Of prime importance for correlated materials are the
local Coulomb interactions between $d$- and $f$-electrons on the
same lattice site since these contributions are largest. This is
due to the extensive overlap  (w.r.t. the Coulomb interaction)
of these localized orbitals
which results in strong correlations.
Moreover, the largest non-local contribution is the
nearest-neighbor density-density interaction which, to leading order in
the number of nearest-neighbor sites, yields only the Hartree
term (see Ref.~\citen{MuHa89} and, also, Ref.~\citen{Wahle98})
 which is already taken into account in the LDA.
To take the local Coulomb interactions into account, one can supplement the LDA
Hamiltonian (\ref{HLDA}) with the local Coulomb matrix approximated
by the (most important) matrix elements
$U_{mm^{\prime }}^{\sigma
\sigma ^{\prime }}$ (Coulomb repulsion and Z-component of
Hund's rule coupling) and ${J_{mm'}}$ (spin-flip terms
of Hund's rule coupling) between the localized electrons (for which we assume $%
i=i_{d}$ and $l=l_{d}$):
\begin{eqnarray} \hat{H} &=&
\hat{H}_{\rm LDA} -\hat{H}_{\rm LDA}^U+
 \frac{1}{2} {\sum_{i=i_d, l= l_d}
\sum_{m\sigma, m^{\prime }\sigma'}
}^{\!\!\!\!\!\prime } \;\; 
{U_{mm^{\prime}}^{\sigma \sigma'}}\hat{n}_{ilm\sigma}\hat{n}_{ilm'\sigma'}
\nonumber \\&&
-\frac{1}{2} {\sum_{i=i_d, l= l_d}}
\sum_{m\sigma, m^{\prime }}^{\;\;\;\;\;\;\;\;\;\prime}
{J_{mm'}^{}}
\hat{c}^\dagger_{ilm\sigma }
\hat{c}^\dagger_{ilm^{\prime}\bar{\sigma}} 
\hat{c}^{\phantom{\dagger}}_{ilm^{\prime}\sigma} 
\hat{c}^{\phantom{\dagger}}_{ilm\bar{\sigma}}.
\label{Hint}
\end{eqnarray}
 Here, the prime on the sum indicates that
at least two of the indices of an operator have to be different, and
$\bar{\sigma}= \downarrow\!(\uparrow)$ for $\sigma = \uparrow\!(\downarrow)$. 
A term $\hat{H}_{{\rm LDA}}^{U}$ is subtracted to avoid
double-counting of those contributions of the local Coulomb
interaction already contained in $\hat{H}_{{\rm LDA}}$. Since
there does not exist a direct microscopic or diagrammatic link
between the model
Hamiltonian approach and LDA it is not possible to express $\hat{H}_{{\rm LDA%
}}^{U}$ rigorously in terms of $U$, $J$ and $\rho $. 
%Guided by the
%observation that the LDA calculates the {\em total energy} of
%isolated atoms rather well, it was argued\cite{Anisimov91} that
%the average energy $E_{{\rm LDA}}^{U}$ corresponding to
%$\hat{H}_{{\rm LDA}}^{U}$ is well approximated by the energy of
%the interaction term in the atomic limit. Hence, in the case of an
%orbital- and spin-independent $U_{mm^{\prime }}^{\sigma \sigma
%^{\prime }}=U$ one may write
A commonly employed
 approximation for $\hat{H}_{{\rm LDA}}^{U}$
assumes the LDA energy $E_{\rm LDA}^U$ of  $\hat{H}_{{\rm LDA}}^{U}$
to be\cite{Anisimov91}
\begin{equation}
E_{\rm LDA}^U=\frac{1}{2}\bar{U}
n_{d}(n_{d}-1)-
\frac{1}{2}{J}\sum_\sigma n_{d\sigma}(n_{d\bar{\sigma}}-1).
\label{ELDAU}
\end{equation}%
%(For the corresponding equation including Hund's rule coupling see Ref. \citen{poter97}). Here, $n_{d}=\sum_{m}n_{il_{d}m}=\sum_{m}\langle \hat{n}%
Here, $n_{d\sigma}=\sum_{m}n_{il_{d}m\sigma}=\sum_{m}\langle \hat{n}%
_{il_{d}m\sigma}\rangle $ is the total number of interacting
electrons per spin, $n_d=\sum_\sigma n_{d\sigma}$,
$\bar{U}$ is the average Coulomb repulsion and ${J}$ the average exchange
or Hund's rule coupling. 
In typical applications we have $U_{mm}^{\uparrow\downarrow}\equiv U$, 
$J_{mm'}^{}\equiv{J}$,
$U_{mm'}^{\sigma\sigma'}= U-J-J\delta_{\sigma\sigma'}$ for $m\ne m'$
(here, the first term $J$ is due to the reduced Coulomb repulsion 
between different orbitals and the second term
$J\delta_{\sigma\sigma'}$ directly arises from  
the Z-component of Hund's rule coupling), and (with the number of interacing orbitals $M$)
$$
\bar U=\frac{U+(M-1)(U-J)+(M-1)(U-2J)}{2M-1}.
$$

Since the one-electron LDA energies can be obtained
from the derivatives of the total energy w.r.t. the occupation
numbers of the corresponding states, the
one-electron energy level for the {\em non-interacting, paramagnetic} states
of (\ref{Hint}%
) is obtained as\cite{Anisimov91}
\begin{equation}
\varepsilon _{il_{d}m}^{0}\equiv \frac{{\rm d}}{{\rm d}n_{il_{d}m}}(E_{{\rm LDA}%
}-E_{{\rm LDA}}^{U})=\varepsilon _{il_{d}m}-\bar{U}(n_{d}-\frac{1}{2})
+\frac{{ J}}{2}(n_d-1)
\label{neweps}
\end{equation}%
where $\varepsilon _{il_{d}m}$ is defined in (\ref{HLDA}) and
$E_{{\rm LDA}}$ is the total energy calculated from $\hat{H}_{{\rm
LDA}}$ (\ref{HLDA}). Furthermore we used $n_{d\sigma}=n_d/2$ in the
paramagnet.

This leads to a new Hamiltonian describing the non-interacting system%
\begin{eqnarray}
\hat{H}^{0}_{{\rm LDA}}&=&\sum_{ilm,jl^{\prime }m^{\prime },\sigma
}\!\!\!(\delta _{ilm,jl^{\prime }m^{\prime }}\varepsilon
_{ilm}^{0}{\hat{n}}_{ilm}^{\sigma
}+t_{ilm,jl^{\prime }m^{\prime }}{\hat{c}}_{ilm}^{\sigma \dagger }{\hat{c}}%
_{jl^{\prime }m^{\prime }}^{\sigma }),
 \end{eqnarray}
where $\varepsilon _{il_{d}m}^{0}$ is given by (\ref{neweps}) for
the interacting orbitals and $\varepsilon _{ilm}^{0}=\varepsilon
_{ilm}$ for the non-interacting orbitals. 
While it is not clear at  present how to systematically
subtract $\hat{H}_{{\rm LDA}}^{U}$ one should note that the
 subtraction of a Hartree-type energy does not substantially 
affect the {\em overall}
behavior of a strongly correlated paramagnetic metal in the
vicinity of a Mott-Hubbard metal-insulator transition (see also
Section \ref{SimpTMO}).

In the following, it is convenient to work in reciprocal space
where the matrix elements of $\hat{H}_{{\rm LDA}}^{0}$, i.e.,
the LDA one-particle energies without the local Coulomb interaction, are given
by 
\begin{eqnarray}
(H^{0}_{{\rm LDA}}({\bf k}))_{qlm,q^{\prime }l^{\prime }m^{\prime
}}\!&\ =\ &\!(H_{{\rm LDA}}({\bf k}))_{qlm,q^{\prime}l^{\prime
}m^{\prime }}\nonumber\\
 & & -\delta _{qlm,q^{\prime }l^{\prime}m^{\prime }}\delta _{ql,q_{d}l_{d}}
\left[\bar{U}(n_{d}-\frac{1}{2})-\frac{J}{2}(n_d-1)\right].
\label{LDAHam}
\end{eqnarray}
Here, $q$ is an index of the atom in the elementary unit cell, $(H_{{\rm LDA}%
}({\bf k}))_{qlm,q^{\prime }l^{\prime }m^{\prime }}$ is the matrix
element of (\ref{HLDA}) in ${\bf k}$-space, and $q_{d}$ denotes
the atoms with
interacting orbitals in the unit cell. The non-interacting part, $\hat{H}_{%
{\rm LDA}}^{0}$, supplemented with the local Coulomb interaction
forms the (approximated) {\em ab initio} Hamiltonian for a
particular material under investigation:
\begin{eqnarray}
\hat{H}& = & 
\hat{H}_{{\rm
LDA}}^{0}+\frac{1}{2}
{\sum_{i=i_{d},l=l_{d}}
\sum_{m\sigma, m^{\prime }\sigma' }}^{\!\!\!\!\!\!\prime }\;\;\;
{U_{mm^{\prime}}^{\sigma \sigma'}}\hat{n}_{ilm\sigma}\hat{n}_{ilm'\sigma'}
\nonumber\\&&
-\frac{1}{2}{\sum_{i=i_{d},l=l_{d}}
\sum_{m\sigma, m^{\prime } }}^{\!\!\!\prime }\;\;\;{J_{mm'}^{}}
\hat{c}^\dagger_{ilm\sigma }
\hat{c}^\dagger_{ilm^{\prime}\bar{\sigma}} 
\hat{c}^{\phantom{\dagger}}_{ilm^{\prime}\sigma} 
\hat{c}^{\phantom{\dagger}}_{ilm\bar{\sigma}}
  \label{HLDAcor}
\end{eqnarray}
To make use of this {\em ab initio} Hamiltonian it is still
necessary to determine the Coulomb interaction $U$. To this end,
one can calculate the LDA ground state energy for different
numbers of interacting electrons $n_{d}$ (''constrained LDA''\cite{Parameters}) and
employ Eq. (\ref{ELDAU}) whose second derivative w.r.t. $n_{d}$
yields $U$. However, one should keep in mind that, while the total
LDA spectrum is rather insensitive to the choice of the basis, the
calculation of $U$ strongly depends on the shape of the orbitals
which are considered to be interacting. E.g., for LaTiO$_3$
at a Wigner Seitz radius of 2.37 a.u. for Ti a LMTO-ASA
calculation\cite{Nekrasov00} using the TB-LMTO-ASA code\cite{LMTO} 
yielded $U=4.2$ eV
in comparison to the value $U=3.2$ eV calculated by ASA-LMTO within
orthogonal representation \cite{Solovyev}.
Thus, an appropriate basis
like LMTO is mandatory and, even so, a significant uncertainty in $U$
remains.

\subsection{Dynamical mean-field theory}
\label{dmft}

The many-body extension of LDA, Eq. (\ref{HLDAcor}), was
proposed by Anisimov {\it et al.}\cite{Anisimov91} in the context of
their LDA+U approach. Within LDA+U the Coulomb interactions of
(\ref{HLDAcor}) are treated within the Hartree-Fock approximation.
Hence,  LDA+U does not contain true many-body physics. While this approach
 is successful in describing long-range ordered, insulating states of
correlated electronic systems it fails to describe strongly correlated
{\em %
paramagnetic} states. To go beyond LDA+U and capture the many-body
nature of the electron-electron interaction, i.e., the frequency
dependence of the self-energy, various approximation schemes have
been proposed and applied
recently\cite{poter97,lichten98,janis98,laegsgaard,wolenski98,zoelfl99}.
One
of the most promising approaches, first implemented by Anisimov et
al.\cite%
{poter97}, is to solve (\ref{HLDAcor}) within DMFT\cite%
{MetzVoll89,MuHa89,Brandt,Vaclav,Georges92,Jarrell92,vollha93,pruschke,georges96} (''LDA+DMFT''). Of all
extensions of LDA only the LDA+DMFT approach is presently able to
describe the physics of {\em strongly} correlated, paramagnetic
metals with well-developed upper and lower Hubbard bands and a
narrow quasiparticle peak at the Fermi level. This characteristic three-peak
structure is a signature of the importance of many-body effects.\cite{Georges92,Jarrell92}

\begin{figure}
\begin{center}\mbox{}
\includegraphics[clip=true,width=12cm]{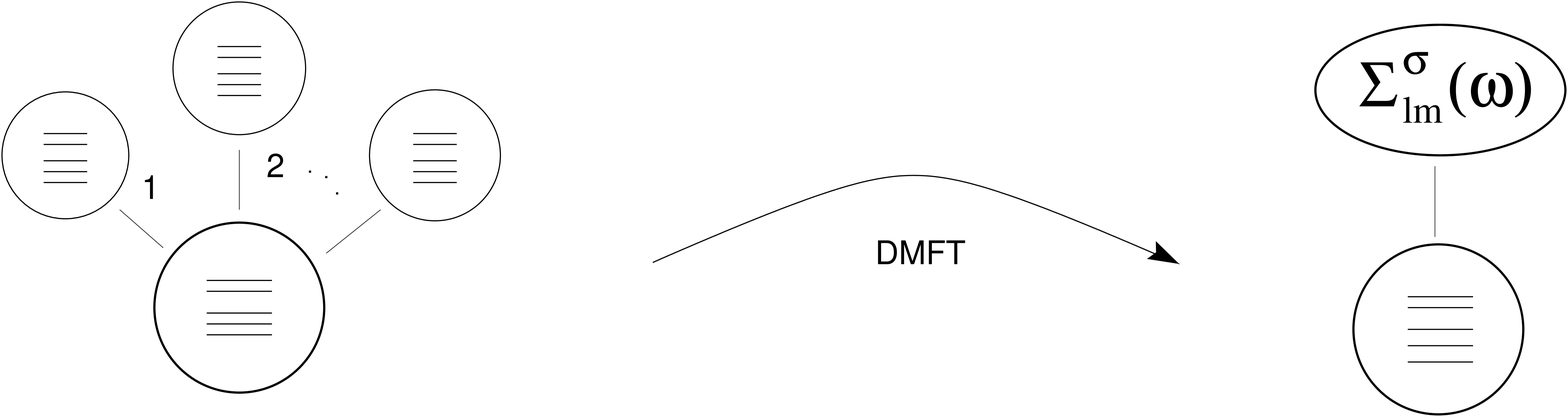}
\end{center}
\caption{If the number of neighboring
lattice sites goes to infinity, the central limit theorem holds
and fluctuations from site-to-site can be neglected. This means 
that the influence of these neighboring sites can be replaced by
a mean influence,  
the dynamical mean-field  described by the self energy 
${\Sigma}_{lm}^{\sigma}(\omega)$.
This DMFT problem is equivalent to the self-consistent solution of the
$\bf k$-integrated Dyson equation (\ref{effAction}) and the multi-band
Anderson impurity model 
Eq.\ (\ref{siam}).}
\label{dmftfig}
\end{figure}

During the last ten years, DMFT has proved to be a successful
approach to
investigate strongly correlated systems with local Coulomb
interactions\cite%
{georges96}. It becomes exact in the limit of high lattice
coordination numbers\cite{MetzVoll89,MuHa89} and preserves the dynamics
of local interactions. Hence, it represents a {\em dynamical}
mean-field approximation. In this non-perturbative approach the
lattice problem is mapped onto an effective single-site problem
(see Fig.~\ref{dmftfig})
which has to be determined self-consistently together
with the ${\bf k}$-integrated Dyson equation connecting the self energy
$%
\Sigma $ and the Green function $G$ at frequency $\omega $:
\begin{eqnarray}
G_{qlm,q^{\prime }l^{\prime }m^{\prime }}(\omega )=\!\frac{1}{V_{B}}\int
{{%
d^{3}}{k}} \!&\left( \left[ \;\omega 1+\mu 1-H_{{\rm LDA}}^{0}({\bf k})
-\Sigma(\omega )\right]^{-1}\right)_{q l m,q^{\prime
}l^{\prime }m^{\prime }} .&  \label{Dyson}
\end{eqnarray}%
Here, $1$ is the unit matrix, $\mu$  the chemical potential, the matrix $H_{{\rm LDA}}^{0}({\bf k})$ is defined in
(\ref{LDAHam}), 
 $\Sigma(\omega)$ denotes the self-energy matrix which is non-zero
only between the interacting orbitals, 
$[...]^{-1}$ implies the inversion of the matrix with elements $n$
(=$%
qlm$), $n^{\prime }$(=$q^{\prime }l^{\prime }m^{\prime }$), and the
integration extends over the Brillouin zone with volume $V_{B}$.

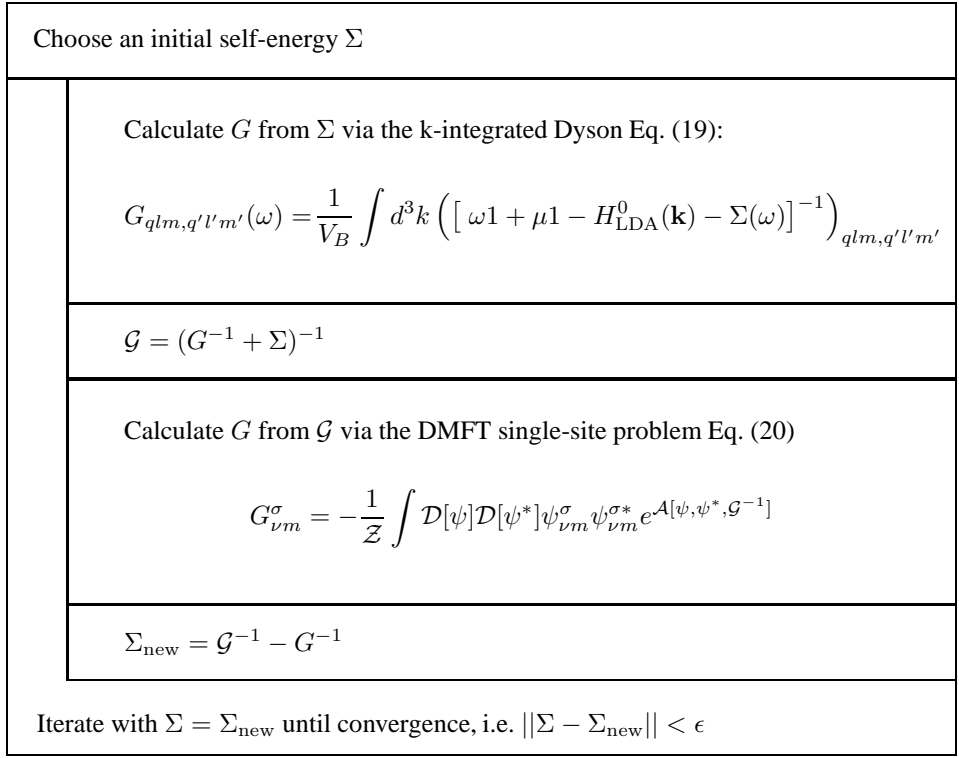
\begin{figure}[tb]
\unitlength=1mm
\special{em:linewidth 0.4pt}
\linethickness{0.4pt}
\begin{picture}(130.00,105.00)
\put(4.00,5.00){\framebox(126.00,90.00)[cc]{}}
\put(4.00,95.00){\framebox(126.00,10.00)[cc]{
\parbox{12cm}{Choose an initial self-energy \protect$\Sigma$}}}

\put(12.00,15.00){\framebox(118.00,80.00)[cc]{}}

\put(7.00,8.00){
\parbox{9.5cm}{Iterate with  $\Sigma =
  \Sigma_{\rm new}$ until convergence, i.e.\ $ ||
  \Sigma -\Sigma_{\rm new}|| < \epsilon $ }}

\put(12.00,65.00){\framebox(118.00,30.00)[cc]
 {\parbox{10.3cm}{Calculate $G$ from $\Sigma$ via the k-integrated Dyson Eq.\ (\ref{Dyson}):\\
\[G_{qlm,q^{\prime }l^{\prime }m^{\prime }}(\omega )=\!\frac{1}{V_{B}}\int
{{
d^{3}}{k}} \left( \left[ \;\omega 1+\mu 1-H_{{\rm LDA}}^{0}({\bf k})
%\right. \right.  
-\Sigma(\omega )\right]^{-1}\right)_{q l m,q^{\prime
}l^{\prime }m^{\prime }} \]
}}}

\put(12.00,55.00){\framebox(118.00,10.00)[cc]
{\parbox{10.3cm}{${\cal G} = ( {G}^{-1}+\Sigma)^{-1}$}}}

\put(12.00,25.00){\framebox(118.00,30.00)[cc]
{\parbox{10.3cm}{
    Calculate $G$ from ${\cal G}$ via the DMFT single-site problem Eq.\ (\ref{siam})\\
\[
G_{\nu m}^{\sigma }=-\frac{1}{{\cal Z}}\int {\cal D}[\psi ]{\cal
D}[\psi
^{\ast }]\psi _{\nu m}^{\sigma \phantom\ast }\psi _{\nu m}^{\sigma \ast
}e^{%
{\cal A}[\psi ,\psi ^{\ast },{\cal G}^{-1}]} 
\]
 }}}

\put(12.00,15.00){\framebox(118.00,10.00)[cc]
{\parbox{10.3cm}{
 ${\Sigma_{\rm new}}= {{\cal G}}^{-1} - G^{-1}$}}}
\end{picture}
\caption{Flow diagram of the DMFT self-consistency cycle.}
\label{DMFTflow}
\end{figure}

The DMFT single-site problem depends on ${\cal
G}(\omega)^{-1}=G(\omega)^{-1}+\Sigma(\omega)$ and is equivalent\cite{Georges92,Jarrell92} to an Anderson impurity
model (the history and the physics of
 this model is summarized by Anderson in Ref.\ \citen{Anderson})
 if its hybridization $\Delta
(\omega )$ satisfies ${\cal G}^{-1}(\omega )=\omega -\int d\omega
^{\prime }\Delta (\omega ^{\prime })/(\omega -\omega ^{\prime })$.
The local one-particle Green function at a Matsubara frequency
$i \omega _{\nu }= i (2\nu +1)\pi /\beta $ ($\beta$: inverse temperature), 
orbital index $m$
($l=l_{d}$, $q=q_{d} $), and spin $\sigma $ is given by the
following functional integral over Grassmann variables $\psi
^{\phantom\ast }$ and $\psi ^{\ast }$:
\begin{equation}
G_{\nu m}^{\sigma }=-\frac{1}{{\cal Z}}\int {\cal D}[\psi ]{\cal
D}[\psi
^{\ast }]\psi _{\nu m}^{\sigma \phantom\ast }\psi _{\nu m}^{\sigma \ast
}e^{%
{\cal A}[\psi ,\psi ^{\ast },{\cal G}^{-1}]}.  \label{siam}
\end{equation}%
Here, ${\cal Z}=\int {\cal D}[\psi ]{\cal
D}[\psi
^{\ast }]\psi _{\nu m}^{\sigma \phantom\ast }\psi _{\nu m}^{\sigma \ast
}\exp(
{\cal A}[\psi ,\psi ^{\ast },{\cal G}^{-1}])$
is the partition function and
the single-site action ${\cal A}$ has the form (the interaction part
of ${\cal A}$ is in terms of the ``imaginary time'' $\tau$, i.e., the
Fourier transform of $\omega_{\nu}$)
% \cite{Held}
\begin{eqnarray}
\lefteqn{{\cal A}[\psi ,\psi ^{\ast },{\cal G}^{-1}]=\sum_{\nu
,\sigma ,m}\psi _{\nu
m}^{\sigma \ast }({\cal G}_{\nu m}^{\sigma })^{-1}\psi _{\nu m}^{\sigma
{%
\phantom\ast }}}\nonumber \\ &&-\frac{1}{2}\!\!\!{\sum_{m\sigma
,m\sigma ^{\prime}}}^{\!\!\!\!\!'}\;\; U_{m m'}^{\sigma \sigma'}
    \int\limits_{0}^{\beta }d\tau \,\psi _{m}^{\sigma \ast }(\tau )\psi
_{m}^{\sigma \phantom\ast }(\tau )\psi _{m^{\prime }}^{\sigma
^{\prime }\ast }(\tau )\psi _{m^{\prime }}^{\sigma ^{\prime }\phantom%
\ast }(\tau )\nonumber\\
&&
+\frac{1}{2} {\sum_{m\sigma
,m}}^{\!\!'}\;\; J_{m m'}^{}
    \int\limits_{0}^{\beta }d\tau \,\psi _{m}^{\sigma \ast }(\tau )\psi
_{m}^{\bar{\sigma}}(\tau )\psi _{m^{\prime }}^{\bar{\sigma}\ast }(\tau )
\psi _{m^{\prime }}^{\sigma}(\tau )\;\;.\label{effAction}
\end{eqnarray}
This single-site problem (\ref{siam}) has to be solved self-consistently together
with the  {\bf k}-integrated Dyson equation (\ref{Dyson}) to obtain the 
DMFT solution 
of a given problem,
see the flow diagram Fig.~\ref{DMFTflow}.

Due to the equivalence of the DMFT single-site problem
 and the Anderson impurity problem a variety
of approximative techniques have been employed to solve the DMFT
equations, such as the iterated perturbation theory (IPT)\cite%
{Georges92,georges96} and the non-crossing approximation (NCA) \cite%
{NCA1,pruschke89,NCA2}, as well as numerical techniques like
quantum Monte Carlo simulations (QMC)\cite{QMC}, exact diagonalization
(ED)\cite%
{caffarel94,georges96}, or numerical renormalization group (NRG)\cite%
{NRG}.  QMC and NCA will be discussed in more
detail in Section \ref{QMC} and \ref{NCA}, respectively.
IPT is non-self-consistent
second-order perturbation theory in $U$ for the Anderson impurity
problem (%
\ref{siam}) at half-filling. It represents an ansatz that also
yields the correct perturbational $U^{2}$-term and the correct
atomic limit for the self-energy off half-filling\cite{Kajueter},
for further details see Refs.~\citen{Kajueter,poter97,lichten98}.
 ED directly diagonalizes the Anderson impurity problem at a limited
number of lattice sites and orbitals.
NRG first replaces the  conduction
band by a discrete set of states at $D \Lambda^{-n}$
($D$: bandwidth; $n=0,...,{\cal N}_s$) and then 
diagonalizes this problem  iteratively with increasing accuracy at low
energies, i.e., with increasing ${\cal N}_s$.
In principle, QMC and ED are exact methods, but they require
an extrapolation, i.e., the discretization of the imaginary time
$\Delta \tau \rightarrow 0$ (QMC) or the number of lattice sites of
the respective impurity model $N_{s}\rightarrow \infty $ (ED), respectively.

In the context of LDA+DMFT we refer
to the computational schemes to solve the DMFT equations
discussed above as LDA+DMFT(X) where X=IPT\cite{poter97}, NCA\cite%
{zoelfl99}, QMC\cite{Nekrasov00} have been investigated in the
case of La$_{1-x}$Sr$_{x}$TiO$_{3}$. The same strategy was
formulated by Lichtenstein and Katsnelson\cite{lichten98} as one
of their LDA++ approaches. Lichtenstein and Katsnelson applied
LDA+DMFT(IPT)\cite{Kats98}, and were the first to use
LDA+DMFT(QMC)\cite{kats99}, to investigate the spectral properties
of iron. Recently, also V$_2$O$_3$\cite{Held01a},
Ca$_{2-x}$Sr$_x$RuO$_4$ \cite{Liebsch00, Anisimov01},
Ni\cite{FeNi}, Fe\cite{FeNi}, Pu\cite{SAVRASOV,SAVRASOV2}, and Ce\cite{Zoelfl01,McMahan01}
have been studied by LDA+DMFT.
Realistic investigations of itinerant ferromagnets (e.g., Ni) have
also recently become possible by combining density functional theory
with multi-band Gutzwiller wave functions.\cite{Gebhard}

\subsection{QMC method to solve DMFT}
\label{QMC}
 The self-consistency cycle of the DMFT (Fig.~\ref{DMFTflow})
requires a method to solve for the
dynamics of the single-site problem of DMFT, i.e., Eq.\ (\ref{siam}).
The QMC algorithm by Hirsch and Fye\cite{QMC} is a well established
method to find a numerically exact solution for the
Anderson impurity model and allows one to calculate
the impurity Green function $G$ at a given ${\cal G}^{-1}$ 
as well as correlation functions. 
In essence, the QMC technique maps the interacting
electron problem Eq.\ (\ref{siam}) onto a sum of non-interacting
problems where the single particle moves in a fluctuating, 
time-dependent field and
evaluates this sum by Monte Carlo sampling,
see the flow diagram Fig.~\ref{QMCflow} for an overview.
To this end, the imaginary time interval $[0,\beta]$ 
of the functional integral Eq.\ (\ref{siam}) is discretized into 
$\Lambda$ steps 
of size $\Delta \tau = \beta/\Lambda$, yielding support points
$\tau_l=l \Delta \tau$ with $l=1 \dots \Lambda$. 
Using this Trotter discretization, 
the integral $\int_0^\beta d\tau$ is transformed to the sum 
$\sum_{l=1}^\Lambda \Delta \tau$ and the exponential terms in Eq. (\ref{siam}) can be 
separated via the Trotter-Suzuki formula for operators 
$\hat{A}$ and $\hat{B}$\cite{suzuki}
\begin{equation}
e^{-\beta (\hat{A}+\hat{B})} = \prod_{l=1}^\Lambda e^{-\Delta \tau \hat{A}} 
e^{-\Delta \tau \hat{B}} + {\cal O}(\Delta \tau),
\end{equation}
which is exact in the limit $\Delta \tau \rightarrow 0$. 
The single site action $\cal A$ of Eq.\ (\ref{effAction}) can now be written
 in the 
discrete,  imaginary time as
\begin{eqnarray}
{\cal A}[
     \psi,\psi^*,{\cal G}^{-1}] &=& \Delta \tau ^2 
     \sum_{ \sigma \,m \, l,l'=
       0}^{\Lambda-1} {\psi_{ m l}^\sigma}^*  
       {{\cal G}_{ m}^\sigma}^{-1} (l\Delta
       \tau - l'\Delta \tau) 
      \psi_{ m l'}^\sigma
    \nonumber \\ && - \frac{1}{2} \Delta \tau {\sum}_{m\sigma,m'\sigma'}^{'} 
    U_{m m'}^{\sigma \sigma'} \sum_{l=0}^{\Lambda-1}
     {\psi_{m l}^\sigma}^*
     {\psi_{m l}^\sigma}
     {\psi_{m' l}^{\sigma'}}^*
     {\psi_{m' l}^{\sigma'}},
\end{eqnarray}
where the first term was Fourier-transformed from Matsubara frequencies
to imaginary time.
In a second step, the $M(2M-1)$ interaction terms
in the single site action 
$\cal A$ are decoupled by introducing a classical auxiliary field 
$s_{l m m'}^{\sigma {\sigma'}}$:
\begin{eqnarray}\label{Hirsch}
 \exp \left\{
   \frac{\Delta \tau}{2} U_{m m'}^{\sigma \sigma'}
     ({\psi_{m l}^\sigma}^*
     {\psi_{m l}^\sigma}-
     {\psi_{m' l}^{\sigma'}}^*
     {\psi_{m' l}^{\sigma'}})^2 \right\} = \qquad \qquad \nonumber \\
\hfill \frac{1}{2} \sum_{s_{l m m'}^{\sigma {\sigma'}}\,  = \pm 1}
\exp \left\{ \Delta \tau \lambda_{l m m'}^{\sigma {\sigma'}}
 s_{l m m'}^{\sigma {\sigma'}} 
     ({\psi_{m l}^\sigma}^*
     {\psi_{m l}^\sigma}-
     {\psi_{m' l}^{\sigma'}}^*
     {\psi_{m' l}^{\sigma'}}) \right\},
\end{eqnarray}
where  $\cosh (\lambda_{l m m'}^{\sigma {\sigma'}}) = 
\exp(\Delta \tau U_{m m'}^{\sigma \sigma'}/2)$ and $M$ is the number 
of interacting orbitals. This  so-called discrete 
Hirsch-Fye-Hubbard-Stratonovich transformation
can be applied to the Coulomb repulsion as well as the Z-component of 
Hund's rule coupling.\cite{QMCproblem} It replaces the interacting 
system by a sum of $\Lambda M (2M-1)$ auxiliary 
fields $s_{l m m'}^{\sigma {\sigma'}}$.
The functional integral can now be solved by a simple Gauss integration
because the Fermion operators only enter quadratically, i.e., for a
given configuration  ${\bf s}=\{ s_{l m m'}^{\sigma {\sigma'}}\}$
of the auxiliary fields the system is non-interacting.
The quantum mechanical problem is then reduced to a matrix problem
\begin{eqnarray}
G_{\tilde{m} l_1 l_2}^{\tilde{\sigma}} = \frac{1}{\cal Z} \frac{1}{2} \sum_l
\sum_{m' \sigma', m'' \sigma''}^{\;\;\;\;\;\;\;\;\;\prime}
\sum_{s_{l m'' m'}^{{\sigma''} {\sigma'}}\,  = \pm 1} 
\left[ (M_{\tilde{m} }^{\tilde{\sigma}\mathbf{s}})^{-1} \right]_{l_1 l_2} 
\prod_{m \sigma} \det \mathbf{M}_{m}^{\sigma \mathbf{s}}
\label{hidimsum}
\end{eqnarray}
with the partition function $\cal Z$, the matrix 
\begin{eqnarray}
\mathbf{M}_{\tilde{m} }^{\tilde{\sigma} \mathbf{s}} = \Delta \tau^2 
[{\mathbf{G}^{\sigma}_{m }}^{-1} + \Sigma_{m}^{\sigma}] 
e^{-\mathbf{\tilde{\lambda}}^{\sigma \mathbf{s}}_{m}} + \mathbf{1} - 
e^{-\mathbf{\tilde{\lambda}}^{\sigma \mathbf{s}}_{m}} 
\label{defM} 
\end{eqnarray}
and the elements of the matrix $\mathbf{\tilde{\lambda}}^{\sigma \mathbf{s}}_{m}$
\begin{eqnarray}
\tilde{\lambda}^{\sigma \mathbf{s}}_{m l l'}= -\delta_{l l'} 
\sum_{m' \sigma'} \lambda_{m m'}^{\sigma \sigma'} 
\tilde{\sigma}_{m m'}^{\sigma \sigma'} s_{l m m'}^{\sigma \sigma'}.
\end{eqnarray}
Here $\tilde{\sigma}_{m m'}^{\sigma \sigma'} = 2 \Theta(\sigma' - \sigma + 
\delta_{\sigma \sigma'}[m'-m]-1)$ changes sign if $(m \sigma)$ and 
$(m' \sigma')$ are exchanged. For more details, e.g., for a derivation of 
 Eq.\ (\ref{defM}) for the matrix $\bf M$,
see Refs.~\citen{QMC,georges96}.

Since the sum in Eq. (\ref{hidimsum}) consists of $2^{\Lambda M(2M-1)}$ 
addends, 
a complete summation for large $\Lambda$ is computationally impossible.
Therefore the Monte Carlo method, which is often an efficient
way  to calculate  
high-dimensional sums and integrals,
 is employed for importance sampling 
of Eq. (\ref{hidimsum}).  In this method, the integrand $F(x)$ is 
split up into a normalized probability distribution $P$ and the remaining 
term $O$:
\begin{equation}
\int dx F(x) = \int dx \,O(x) \,P(x) \equiv \langle O\rangle_P
\end{equation}
with
\begin{equation}
\int dx P(x) = 1 \qquad \text{and} \qquad P(x) \ge 0.
\end{equation}
In statistical physics, the Boltzmann distribution is often a good choice 
for the function $P$:
\begin{equation}
P(x) = \frac{1}{\cal Z} \exp(-\beta E(x)).
\end{equation}
For the sum of Eq.\ (\ref{hidimsum}), this probability distribution
 translates to
\begin{equation}
P(\mathbf{s}) = \frac{1}{\cal Z} \prod_{m \sigma} \det 
\mathbf{M}_{m }^{\sigma \mathbf{s}}
\end{equation}
with the remaining term
\begin{equation}
O(\mathbf{s})_{\tilde{m} l_1 l_2}^{\tilde{\sigma}} = 
\left[ (M_{\tilde{m}}^{\tilde{\sigma}\mathbf{s}})^{-1} \right]_{l_1 l_2}.
\label{curG}
\end{equation}

\begin{figure}[t]
\unitlength=1mm
\special{em:linewidth 0.4pt}
\linethickness{0.4pt}
\begin{picture}(130.00,85.00)

%\put(0.00,75.00){\framebox(130.00,10.00)[cc]
%{\parbox{12cm}{Set $\;\;\;G^\sigma_{m l_1 l_2} = 0$}}}

\put(0.00,80.00){\framebox(130.00,10.00)[cc]
{\parbox{12cm}{Choose random auxiliary field configuration $\mathbf{s}=\{s_{l m m'}^{\sigma {\sigma'}}\}$}}}

\put(0.00,50.00){\framebox(130.00,30.00)[cc]
{\parbox{12cm}{
Calculate the current Green function ${\mathbf G}_{\rm cur}$ from Eq.\ (\ref{curG}) \[
({G_{\rm cur}})_{\tilde{m} l_1 l_2}^{\tilde{\sigma}} = 
\left[ (M_{\tilde{m}}^{\tilde{\sigma}\mathbf{s}})^{-1} \right]_{l_1 l_2}
\] 
with  ${\mathbf M}$ from Eq.\ (\ref{defM}) and
the input  ${{\cal G}^{\sigma}_{m }}(\omega_{\nu})^{-1}={{G}^{\sigma}_{m }}(\omega_{\nu})^{-1}
+ \Sigma_{m}^{\sigma}(\omega_{\nu})$.
}}}

\put(0.00,30.00){\framebox(130.00,20.00)[cc]{}}

\put(5.00,45.00){
\parbox{13cm}{Do NWU times (warm up sweeps)}}

\put(10.00,30.00){\framebox(120.00,10.00)[cc]
{\parbox{11cm}{ \em MC-sweep $({\mathbf G}_{\rm cur}, {\mathbf s})$}}}

\put(0.00,0.00){\framebox(130.00,30.00)[cc]{}}

\put(5.00,25.00){
\parbox{13cm}{Do NMC times (measurement sweeps)}}

\put(10.00,10.00){\framebox(120.00,10.00)[cc]
{\parbox{11cm}{ \em MC-sweep  $({\mathbf G}_{\rm cur}, {\mathbf s})$}}}

\put(10.00,0.00){\framebox(120.00,10.00)[cc]
{\parbox{11cm}{${\mathbf G} = {\mathbf G} + {\mathbf G}_{\rm cur}/ {\rm NMC}$}}}

%\put(10,25){\line(6,-1){60}}
%\put(130,25){\line(-6,-1){60}}
%\put(70,15){\line(0,-1){15}}
%\put(10,25){\line(0,-1){10}}
\end{picture}
\caption
{Flow diagram of the QMC algorithm to calculate the Green
function matrix ${\mathbf G}$ using the
procedure {\em MC-sweep} of Fig.~\ref{mcsweep}.\label{QMCflow}}
\begin{picture}(130.00,70.00)

\put(0.00,0.00){\framebox(130.00,64.00)[cc]{}}

\put(5.00,55.00){
\parbox{12cm}{Choose $M (2M-1) \Lambda $ times a  set
($l \, m \, m' \, \sigma \, \sigma'$),\\ define ${\mathbf s}_{\rm new}$
to be ${\mathbf s}$ except for the element
$s_{l m m'}^{\sigma {\sigma'}}$ which has opposite sign.}}

\put(10.00,25.00){\framebox(120.00,25.00)[cc]
{\parbox{11cm}{Calculate flip probability
$P_{{\mathbf s}\to {\mathbf s}_{\rm new}}  = {\rm min} \{1,P({\mathbf s}_{\rm new})/P({\mathbf s})\}$ with
\[
P({\mathbf s}_{\rm new})/P({\mathbf s})=\prod_{m \sigma} \det \mathbf{M}_{m}^{\sigma \mathbf{s}_{\rm new}}/\prod_{m \sigma} \det \mathbf{M}_{m}^{\sigma \mathbf{s}}
\]
 and ${\mathbf M}$ from Eq.\ (\ref{defM}).}}}

\put(10,25){\line(5,-1){75}}
\put(130,25){\line(-3,-1){45}}
\put(85,10){\line(0,-1){10}}
\put(10,25){\line(0,-1){15}}

\put(50.00,21.00){
\parbox{6cm}{{\rm Random number} $\in {\mathrm (0,1)} <P_{{\mathbf s}\to {\mathbf s}_{\rm new}}$  ?}}
\put(20.00,12.00){
\parbox{3cm}{{\rm yes}}}
\put(110.00,12.00){
\parbox{3cm}{{\rm no}}}

\put(10.00,0.00){\framebox(120.00,10.00)[cc]{}}
\put(11.00,5.00){
\parbox{12cm}{${\mathbf s}={\mathbf s}_{\rm new}$; recalculate  ${\mathbf G}_{\rm cur}$ according to  Eq.\ (\ref{curG}). \hspace{.5cm} Keep ${\mathbf s}$
}}

\end{picture}
\caption
{
Procedure {\em MC-sweep} using the Metropolis\cite{Metropolis} 
rule to  change the sign of $s_{l m m'}^{\sigma {\sigma'}}$. The recalculation of  ${\mathbf G}_{\rm cur}$, 
i.e., the matrix $\mathbf{M}$ of Eq.\ (\ref{defM}), simplifies
to ${\cal O}(\Lambda^2)$ operations if only one  
$s_{l m m'}^{\sigma {\sigma'}}$ changes sign.\cite{QMC,georges96}\label{mcsweep}}
\end{figure}

Instead of summing over all possible configurations, the Monte Carlo simulation
generates  configurations 
$x_i$ with respect to the probability distribution $P(x)$ and 
averages 
the observable $O(x)$ over these $x_i$. Therefore the relevant 
parts of the phase space with a large Boltzmann weight are taken into 
account to a greater extent than the ones with a small weight, coining the 
name importance sampling for this method.
With the central limit theorem one gets for $\cal N$ statistically independent 
addends the estimate
\begin{equation}
\langle O\rangle_P = \frac{1}{\cal N} \sum^{\cal N}_{\underset{x_i \in P(x)}{i=1}}
O(x_i) \pm \frac{1}{\sqrt{\cal N}} \sqrt{\langle O^2\rangle_P - \langle O\rangle_P^2}.
\end{equation}
Here, the error and with it the number of needed addends $\cal N$ is nearly 
independent of the dimension of the integral. The computational effort for 
the Monte Carlo method is therefore only rising polynomially with the 
dimension of the integral and not exponentially as in a normal integration.
Using a Markov process and single spin-flips in the auxiliary fields, the 
computational cost of the algorithm in leading order of $\Lambda$ is
\begin{equation}
2 {a} M (2M-1) \Lambda^3 \times \text{number of MC-sweeps},
\end{equation}
where $a$ is the acceptance rate for a single spin-flip.

The advantage of the QMC method (for the algorithm see the flow diagram Fig.~\ref{QMCflow})
is that it is (numerically) exact. It allows one to calculate
the one-particle Green function as well as two-particle (or higher) Green 
functions. On present workstations the QMC approach
is able to deal with up to 
seven {\em interacting} orbitals and temperatures 
above about room temperature.
Very
low temperatures are not accessible because the numerical effort
grows like $\Lambda^3\propto 1/T^3$ . Since the QMC approach calculates
$G(\tau)$ or $G(i \omega_n)$ with a statistical error, it also requires the
maximum entropy method\cite{MEM} to obtain the Green function 
$G(\omega)$ at real (physical) frequencies $\omega$.

\subsection{NCA method to solve DMFT}
\label{NCA}
The NCA approach is a
resolvent perturbation theory in the hybridization parameter
$\Delta (\omega )$ of the effective
Anderson impurity problem\cite{NCA1}. Thus, it is
reliable if the Coulomb interaction $U$ is large compared to the
band-width and also offers a computationally inexpensive
approach to  check the
general spectral features 
in other situations.

To see how the NCA can be adapted for the DMFT, let us rewrite Eq.\ (\ref{Dyson})
as
\begin{equation}\label{Dyson2}
G_{\sigma}(z)
=\frac{1}{N_k} \sum_{{\bf k}} \big[{z-H_{LDA}^0({\bf k})-\Sigma(z)}\big]^{-1}
\end{equation}
where $z=\omega+i 0^+ + \mu$.
Again, $H_{LDA}^0({\bf k})$, $\Sigma(z)$ and hence $G^0_{\sigma}(\zeta)$
and $G_{\sigma}(z)$ are matrices in orbital space. Note
that $\Sigma(z)$ has nonzero entries for the correlated orbitals only.

On quite general grounds, Eq.\ (\ref{Dyson2}) can be cast into the form
\begin{equation}
G_{\sigma}(z)=
\frac{1}{z-E^0-\Sigma_{\sigma}(z)-\Delta_{\sigma}(z)}
\end{equation}
where
\begin{equation}\label{h0loc}
E^0=\frac{1}{N_k}\sum\limits_{\bf k}H_{LDA}^0({\bf k})
\end{equation}
with the number of ${\bf k}$ points $N_k$ and
\begin{equation}
\lim_{\omega\rightarrow\pm\infty}\Re e\{\Delta_{\sigma}(\omega+i\delta)\}
=0\;\;.
\end{equation}

Given the the matrix $E^0$, the Coulomb matrix $U$ and
the hybridization matrix $\Delta_{\sigma}(z)$, we are now in a position to
set up a resolvent perturbation theory with respect to $\Delta_{\sigma}(z)$.
To this end, we first have to diagonalize the local Hamiltonian
\begin{equation}
\begin{array}{l@{\,=\,}l}
\displaystyle
H_{\rm local}
&
\begin{array}[t]{l}\displaystyle
\sum\limits_\sigma
\sum\limits_{qml}\sum\limits_{q'm'l'}c^\dagger_{qlm\sigma}
E^0_{qlm,q'l'm'}
c^{\phantom{^\dagger}}_{qlm\sigma}\\[7.5mm]
\displaystyle
+\frac{1}{2}\sum\limits_{m\sigma}\sum\limits_{m'\sigma'}U_{mm'}^{\sigma\sigma'}
n_{q_dl_dm\sigma}n_{q_dl_dm'\sigma'}\\[7.5mm]
\displaystyle
-\frac{1}{2}\sum\limits_{m\sigma}\sum\limits_{m'}J_{mm'}^{}
c^\dagger_{q_dl_dm\sigma}c^\dagger_{q_dl_dm'\bar{\sigma}}
c^{\phantom{\dagger}}_{q_dl_dm'\sigma}c^{\phantom{\dagger}}_{q_dl_dm\bar{\sigma}}\\[7.5mm]
\end{array}\\
&
\displaystyle\sum\limits_\alpha E_\alpha|\alpha\rangle\langle\alpha|
\end{array}
\end{equation}
with local eigenstates $|\alpha\rangle$ and energies $E_\alpha$.
In contrast to the QMC, this approach allows one to take into account the full
Coulomb matrix plus spin-orbit coupling. 

With the states $|\alpha\rangle$ defined above, the fermionic operators with
quantum numbers $\kappa=(q,l,m)$ are expressed as
\begin{equation}
\begin{array}{l@{\,=\,}l}
\displaystyle
c^\dagger_{\kappa\sigma} 
& 
\displaystyle
\sum\limits_{\alpha,\beta}
\left(D^{\kappa\sigma}_{\beta\alpha}\right)^* |\alpha\rangle \langle \beta|\:,\\[5mm]
\displaystyle
c^{\phantom{dagger}}_{\kappa\sigma}
&
\displaystyle
\sum\limits_{\alpha,\beta}
D^{\kappa\sigma}_{\alpha\beta}  |\alpha\rangle \langle \beta|\;\;.
\end{array}
\end{equation}
The key quantity for the resolvent perturbation theory is the resolvent
$R(z)$, which obeys a Dyson equation \cite{NCA1}
\begin{equation}
R(z)=R^0(z)
+R^0(z)S(z)R(z)\;\;,
\end{equation} 
where $R_{\alpha\beta}^0(z)=1/(z-E_\alpha)\delta_{\alpha\beta}$
and $S_{\alpha\beta}(z)$ denotes the self-energy for the local states due to the
coupling to the environment through $\Delta(z)$. 

The self-energy $S_{\alpha\beta}(z)$ can be expressed as power series in the
hybridization $\Delta(z)$ \cite{NCA1}. Retaining only
the lowest-, i.e.\ $2^{nd}$-order terms leads to a set of self-consistent
integral equations
\begin{equation}\label{Snca}
\begin{array}{c}\displaystyle
S_{\alpha\beta}(z)=\begin{array}[t]{l}
\displaystyle
\sum\limits_\sigma\sum\limits_{\kappa\kappa'}
\sum\limits_{\alpha'\beta'}
\int\frac{d\varepsilon}{\pi}\,f(\varepsilon)\,
\left(D^{\kappa\sigma}_{\alpha'\alpha}\right)^*
\Gamma^{\kappa\kappa'}_\sigma(\varepsilon)
R_{\alpha'\beta'}(z+\varepsilon)
D^{\kappa'\sigma}_{\beta'\beta}\;\\[5mm]
\displaystyle
+\sum\limits_\sigma\sum\limits_{\kappa\kappa'}
\sum\limits_{\alpha'\beta'}
\int\frac{d\varepsilon}{\pi}\,(1-f(\varepsilon))\,
D^{\kappa\sigma}_{\alpha'\alpha}
\Gamma^{\kappa\kappa'}_\sigma(\varepsilon)
R_{\alpha'\beta'}(z-\varepsilon)
\left(D^{\kappa'\sigma}_{\beta'\beta}\right)^*
\end{array}
\end{array}
\end{equation}
to determine $S_{\alpha\beta}(z)$, where $f(\varepsilon)$ denotes Fermi's function 
and $\Gamma(\varepsilon)=-\Im m\left\{\Delta(\varepsilon+i0^+)\right\}$.
The set of equations (\ref{Snca}) are in the
literature referred to as non-crossing
approximation (NCA), because, when viewed in terms of diagrams, they contain
no crossing of band-electron lines.
In order to close the cycle for the DMFT, we still have to calculate the true
local Green function $G_\sigma(z)$. This, however, can be done
within the same approximation with the result
\begin{equation}\label{GfNCA}
G^{\kappa\kappa'}_{\sigma}(i\omega) = \frac{1}{Z_{\rm local}}
\sum\limits_{\alpha,\alpha'}\sum\limits_{\nu,\nu'}
D^{\kappa\sigma}_{\alpha\alpha'}\left(D^{\kappa'\sigma}_{\nu\nu'}\right)^*
\oint\frac{dz e^{-\beta z}}{2\pi i}
R_{\alpha\nu}(z)R_{\alpha'\nu'}(z+i\omega)\;\;.
\end{equation}
Here, $\displaystyle Z_{\rm local}=
\sum\limits_\alpha\oint\frac{dze^{-\beta z}}{2\pi i} 
R_{\alpha\alpha}(z)$ 
denotes the local partition function and $\beta$ is the inverse temperature.

Like any other technique, the NCA has its merits and disadvantages. As a
self-consistent resummation of diagrams it constitutes a conserving approximation
to the Anderson impurity model. Furthermore, it is a (computationally)
fast method to
obtain dynamical results for this model and thus also within DMFT. However,
the NCA is known to violate Fermi liquid properties at temperatures much lower
than the smallest energy scale of the problem and whenever charge excitations
become dominant\cite{MH,NCA2}. Hence, in some parameter ranges it fails
in the most dramatic way and must therefore be applied with considerable
care\cite{NCA2}.

\subsection{Simplifications for transition metal oxides with well separated
  $e_{g}$- and $t_{2g}$-bands}
\label{SimpTMO}

Many transition metal oxides are cubic perovskites, with
only a slight distortion of the cubic crystal structure. In these
systems the transition metal $d$-orbitals lead to strong Coulomb
interactions between the electrons. The cubic crystal-field of the
oxygen causes the $d$-orbitals
to split into three degenerate $t_{2g}$- and two degenerate $e_{g}$%
-orbitals. This splitting is often so strong that the $t_{2g}$- or $e_{g}$%
-bands at the Fermi energy are rather well separated from all
other bands. In this situation the low-energy physics is well
described by taking only the degenerate 
bands at the Fermi energy into
account. Without symmetry breaking, the
Green function and the self-energy of these bands 
remain degenerate, i.e., $%
G_{qlm,q^{\prime }l^{\prime }m^{\prime }}(z)=G(z)\delta
_{qlm,q^{\prime }l^{\prime }m^{\prime }}$ and $\Sigma
_{qlm,q^{\prime }l^{\prime }m^{\prime }}(z)=\Sigma (z)\delta
_{qlm,q^{\prime }l^{\prime }m^{\prime }}$ for $l=l_{d} $ and
$q=q_{d}$ (where $l_{d}$ and $q_{d}$ denote the electrons in the
interacting band at the Fermi energy). Downfolding to a basis with
these
degenerate $q_{d}$-$l_{d}$-bands results in an effective Hamiltonian $H_{%
{\rm LDA}}^{0\;{\rm eff}}$ (where indices $l=l_{d}$ and $q=q_{d}$
are suppressed) \begin{eqnarray} G_{m m'}(\omega)
&=&\frac{1}{{V_{B}}} \int {\rm d}^3 k \;
\left([\omega 1+\mu 1-H^{0 \; \rm
eff}_{{\rm LDA}}({\bf k})-\Sigma(\omega) ]^{-1}\right)_{m m^{\prime}}.
\end{eqnarray}
 Due to the diagonal structure of the self-energy
the degenerate interacting Green function can be expressed via the
 non-interacting Green function $G^{0}(\omega)$:
\begin{eqnarray}
G(\omega)&\!=\!&G^{0}(\omega-\Sigma (\omega))=\int d\epsilon
\frac{N^{0}(\epsilon )}{\omega-\Sigma (\omega)-\epsilon}.
\label{intg}
\end{eqnarray}
Thus, it is possible to use the Hilbert transformation of the
unperturbed
LDA-calculated density of states (DOS) $N^{0}(\epsilon )$, i.e., Eq. (\ref%
{intg}), instead of Eq. (\ref{Dyson}). This simplifies the
calculations considerably. With Eq. (\ref{intg}) also some
conceptual simplifications arise: (i) the subtraction of
$\hat{H}_{{\rm LDA}}^{U}$ in (\ref{intg}) only results in an
(unimportant) shift of the chemical potential and, thus, the exact
form of $\hat{H}_{{\rm LDA}}^{U}$ is irrelevant; (ii) Luttinger's
theorem of Fermi pinning holds, i.e., the interacting DOS at the
Fermi energy is fixed at the value of the non-interacting DOS at
$T=0$ within a Fermi liquid; (iii) as the number of electrons
within the different bands is fixed, the LDA+DMFT approach is
automatically self-consistent.

In this context it should be noted that the approximation Eq.
(\ref{intg}) is justified only if the overlap between the $t_{2g}$
orbitals and the other orbitals is rather weak.

\subsection{Extensions of the LDA+DMFT scheme}
\label{selfconsist}

In the present form of the LDA+DMFT scheme the band-structure input due
to LDA and the inclusion of the electronic correlations by DMFT are
performed as successive steps without subsequent feedback. 
In general, the DMFT solution will result in a change of the
occupation of the different bands involved. This changes the
electron density $\rho ({\bf r})$ and, thus, results in a new
LDA-Hamiltonian $\hat{H}_{\rm LDA}$ (\ref{HLDA}) since $\hat{H}_{\rm LDA}$
depends on $\rho ({\bf r})$. At the same time also the Coulomb
interaction $U$ changes and needs to be determined by a new
constrained LDA calculation.
In  a {\it self-consistent} LDA+DMFT scheme, $H_{\rm LDA}$
and $U$ would
define a new Hamiltonian (\ref{HLDAcor}) which again needs to be
solved within DMFT, etc.,
 until convergence is reached:
\begin{equation}
%\rho ({\bf r}) \longrightarrow  H_{\rm LDA}, U \;
%\stackrel{DMFT}{\longrightarrow}\; n_{ilm} \longrightarrow \rho ({\bf
%r}) ...
\setlength{\unitlength}{3947sp}%
\begin{picture}(3762,497)(901,70)
\thinlines
\put(4651,164){\line( 0,-1){225}}
\put(4651,-61){\line(-1, 0){2550}}
\put(2101,-61){\vector( 0, 1){225}}
\put(3926,314){\vector( 1, 0){600}}
\put(1001,314){\vector( 1, 0){600}}
\put(2551,314){\vector( 1, 0){900}}
\put(901,239){\makebox(0,0)[lb]{$\!\!\!\! \!\!\! \!\!\! \rho({\bf r})
\;\;\;$}}
\put(2701,389){\makebox(0,0)[lb]{DMFT}}
\put(1876,239){\makebox(0,0)[lb]{$\!\!\!\!\!\!H_{\rm LDA}$, $U\;\;\;$ }}
\put(3526,239){\makebox(0,0)[lb]{$n_{ilm}\;\;\;$}}
\put(4576,239){\makebox(0,0)[lb]{$\rho({\bf r})$}}
\end{picture}
\vspace{1.5em}
\label{eq:sc}
\end{equation}
\vspace{.5em}

\noindent Without Coulomb interaction ($U=0$) this scheme reduces to the
self-consistent solution of the Kohn-Sham equations. A self-consistency
scheme similar to Eq. (\ref{eq:sc}) was employed by Savrasov and
Kotliar\cite{SAVRASOV2}
in their calculation of Pu. An {\em ab initio}
DMFT scheme formulated directly in the continuum was recently proposed
by Chitra and Kotliar.\cite{contDMFT}

\section{Comparison of different methods to solve DMFT: 
          the model system La$_{1-x}$Sr$_{x}$TiO$_{3}$}
\label{LaTiO3} 

The stoichiometric compound LaTiO$_{3}$
is a cubic perovskite with a small orthorhombic distortion
($\angle ~Ti-O-Ti~\approx
~155^{\circ }$)\cite{maclean79} and is an antiferromagnetic insulator\cite%
{eitel86} below $T_{N}=125$~K\cite{gopel}. Above $T_{N}$, or at
low Sr-doping $\ x$, and neglecting the small orthorhombic
distortion (i.e., considering a cubic structure with the same
volume), LaTiO$_{3}$ is a
strongly correlated, but otherwise simple paramagnet with only {\em one} 3$d$%
-electron on the trivalent Ti sites. This makes the system a
perfect trial candidate for the LDA+DMFT approach.

\begin{figure}[tbp]
 \centering
\includegraphics[clip=true,width=8cm]
{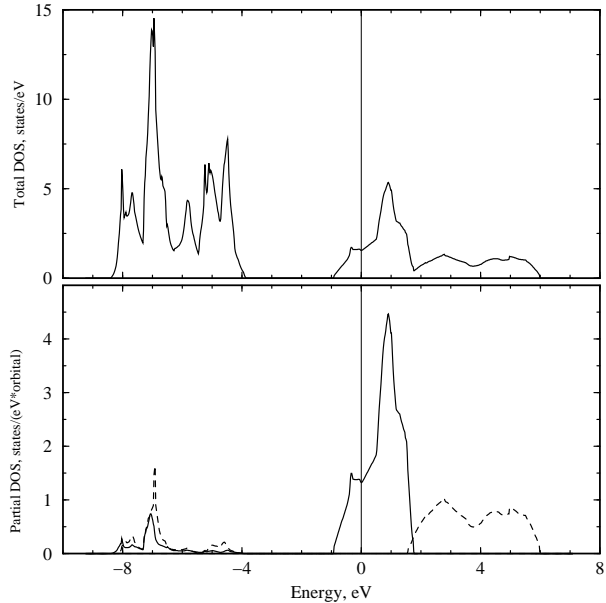}
%\vspace{8cm}

 \caption{Densities of states of LaTiO$_3$ calculated with LDA-LMTO.
Upper
figure: total DOS; lower figure: partial $t_{2g}$ (solid lines)
and $e_g$ (dashed lines) DOS [reproduced from Ref.\protect
\citen{Nekrasov00}].} \label{ldados}
\end{figure}

\begin{figure}[tbp]
\centering
\includegraphics[clip=true,width=8cm]{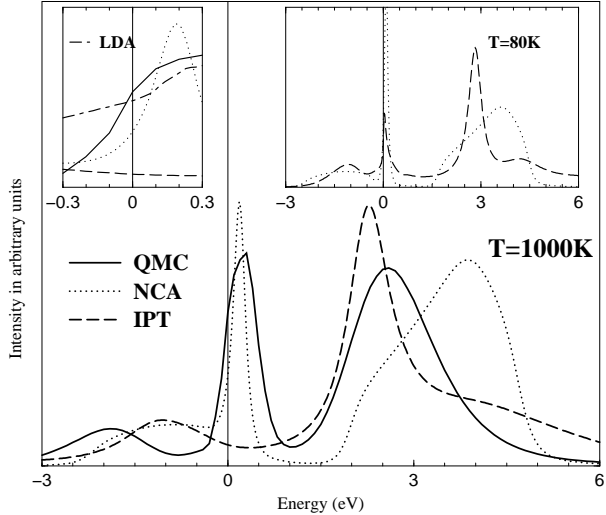}
%\vspace{7cm}

\caption{Spectrum  of La$_{0.94}$Sr$_{0.06}$TiO$_{3}$
 as calculated by LDA+DMFT(X) at $T=0.1$~eV ($\approx 1000$~K)
and $U=4$~eV employing the approximations X=IPT, NCA, and
numerically exact QMC. Inset left: Behavior at the Fermi level
including the LDA DOS. Inset right:  X=IPT and NCA spectra at
$T=80$~K
 [reproduced from Ref.\protect \citen{Nekrasov00}].}
\label{dmft_latio}
\end{figure}

The LDA band-structure calculation 
for undoped  (cubic) LaTiO$_{3}$
yields the DOS shown in Fig. \ref{ldados} which is typical for
early transition metals. The oxygen bands, ranging from $-8.2$~eV
to $-4.0$~eV, are filled such that Ti is three-valent. Due to the
crystal-field splitting, the Ti 3$d$-bands separates into two
empty $e_{g}$-bands and three degenerate $t_{2g}$-bands. Since the
$t_{2g}$-bands at the Fermi energy are well separated also from the
other bands we employ the approximation introduced in section 2.5
which allows us to work with the LDA DOS [Eq.\ (\ref{intg})]
instead of the full one-particle Hamiltonian $H_{\rm LDA}^0$ of
[Eq.\ (\ref{Dyson})].
In the LDA+DMFT calculation, Sr-doping $x$ is taken into account
by adjusting the
chemical potential to yield $n=1-x=0.94$ electrons within the $t_{2g}$%
-bands, neglecting effects disorder and the $x$-dependence of the LDA 
DOS (note, that Sr and Ti have a very similar band structure within LDA). 
There is some uncertainty in the LDA-calculated Coulomb
interaction parameter $U$ $\sim 4-5$ eV (for a discussion see Ref.~\citen%
{Nekrasov00}) which is here assumed to be spin- and
orbital-independent.
In Fig.~\ref{dmft_latio}, results for the spectrum of
La$_{0.94}$Sr$_{0.06}$%
TiO$_{3}$ as calculated by LDA+DMFT(IPT, NCA, QMC) for the same LDA DOS
at $%
T\approx 1000$~K and $U=4$~eV are compared\cite{Nekrasov00}. In
Ref.~\citen{Nekrasov00} the formerly presented IPT\cite{poter97} and
NCA\cite{zoelfl99} spectra were recalculated to allow for a
comparison at exactly the same parameters. All three
methods yield the
typical features of strongly correlated metallic paramagnets: a
lower Hubbard band, a quasi-particle peak (note that IPT produces
a quasi-particle peak only below about 250K which is therefore not
seen here), and an upper Hubbard band. By contrast, within LDA the
correlation-induced Hubbard bands are missing and only a broad
central quasi-particle band (actually a one-particle peak) is
obtained (Fig.~\ref{ldados}).

While the results of the three evaluation techniques of the DMFT
equations (the approximations IPT, NCA and the numerically exact
method QMC) agree on a qualitative level, Fig.~\ref{dmft_latio}
reveals considerable quantitative differences. In particular, the
IPT quasi-particle peak found at low temperatures (see right inset
of Fig.~\ref{dmft_latio}) is too narrow such
that it disappears already at about 250~K and is, thus, not present at $%
T\approx 1000$~K. A similarly narrow IPT quasi-particle peak was
found in a
three-band model study with Bethe-DOS by Kajueter and Kotliar\cite{Kajueter}%
. Besides underestimating the Kondo temperature, IPT also produces
notable deviations in the shape of the upper Hubbard band.
Although NCA comes off much better than IPT it still
underestimates the width of the quasiparticle peak by a factor of
two. Furthermore, the position of the quasi-particle
peak is too close to the lower Hubbard band. In the left inset of Fig.~\ref%
{dmft_latio}, the spectra at the Fermi level are shown. At the
Fermi level, where at sufficiently low temperatures the
interacting DOS should be pinned at the non-interacting value, the
NCA yields a spectral function which is almost by a factor  of two too
small. The shortcomings of the NCA-results, with a too small low-energy scale
and too much broadened Hubbard bands for multi-band systems,  are well understood
and related to the neglect of exchange type diagrams.\cite{wernerH}
Similarly, the
deficiencies of the IPT-results are not entirely surprising in
view of the semi-phenomenological nature of this approximation,
especially for a system off half filling.

This comparison shows
that the choice of the {\em method} used to solve the DMFT equations is
indeed {\em important}, and that, at least for the present system, 
the approximations IPT and NCA differ quantitatively from the
numerically exact QMC. Nevertheless, the NCA gives a rather good account
of the qualitative spectral features and, because it is fast and can
often be applied to comparatively low temperatures, can yield
 an overview of the physics to be expected.

\begin{figure}[tbp]
 \centering
\includegraphics[clip=true,width=8cm]
 {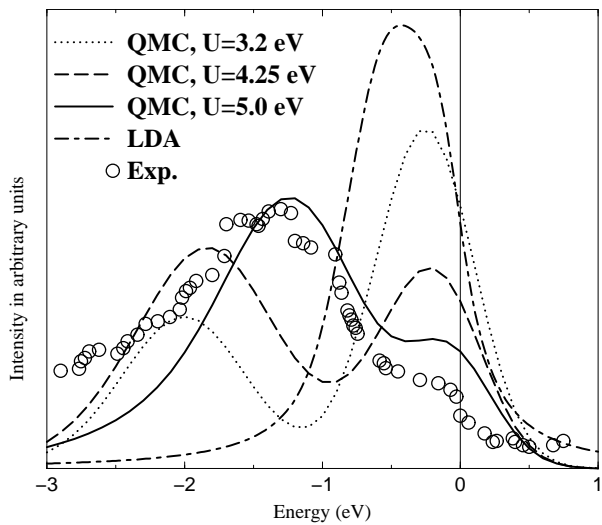}
%\vspace{7cm}

\caption{Comparison of the experimental photoemission
spectrum\protect\cite{fujimori}, the  LDA result, and the
LDA+DMFT(QMC) calculation for
 La$_{0.94}$Sr$_{0.06}$TiO$_{3}$ (i.e., 6\% hole doping) and different
Coulomb interaction $U=3.2$, $4.25$, and $5$~eV [reproduced from
Ref.\protect\citen{Nekrasov00}]. \label{explatio}}
\end{figure}

Photoemission spectra provide a direct experimental tool to study
the electronic structure and spectral properties of electronically
correlated materials. A comparison of LDA+DMFT(QMC) at
1000~K\cite{Note} with the
experimental photoemission spectrum\cite{fujimori} of La$_{0.94}$Sr$_{0.06}$%
TiO$_{3}$ is presented in Fig~\ref{explatio}. To take into account
the uncertainty in $U$\cite{Nekrasov00}, we present results for
$U=3.2$, $4.25$ and $5$ eV. All spectra are multiplied with the
Fermi step function and are Gauss-broadened with a broadening
parameter of 0.3~eV to simulate the experimental
resolution\cite{fujimori}. LDA band structure calculations, the
results of which are also presented in Fig.~\ref{explatio},
clearly fail to reproduce the broad band observed in the
experiment at 1-2~eV below the Fermi energy\cite{fujimori}. Taking
the correlations between the electrons into account, this lower
band is easily identified as the lower Hubbard band whose spectral
weight originates from the quasi-particle band at the Fermi energy
and which increases with $U$. The best agreement with experiment
concerning the relative intensities of the Hubbard band and the
quasi-particle peak and, also, the position of the Hubbard band is
found for $U=5$ eV. The value $U=5$ eV is still compatible with
the {\em ab initio} calculation of this parameter within
LDA\cite{Nekrasov00}. One should also bear in mind that
photoemission experiments are sensitive to surface properties. Due
to the reduced coordination number at the surface the bandwidth is
likely to be smaller, and the Coulomb interaction less screened,
i.e., larger. Both effects make the system more correlated and,
thus, might also explain why better agreement is found for $U=5$
eV. Besides that, also the polycrystalline nature of the sample,
as well as spin and orbital\cite{Keimer} fluctuation not taken
into account in the LDA+DMFT approach, will lead to a further
reduction of the quasi-particle weight.

\section{Mott-Hubbard metal-insulator transition in V$_2$O$_3$}
\label{v2o3}

\begin{figure}[tbp]
 \centering
\includegraphics[clip=true,width=9cm]{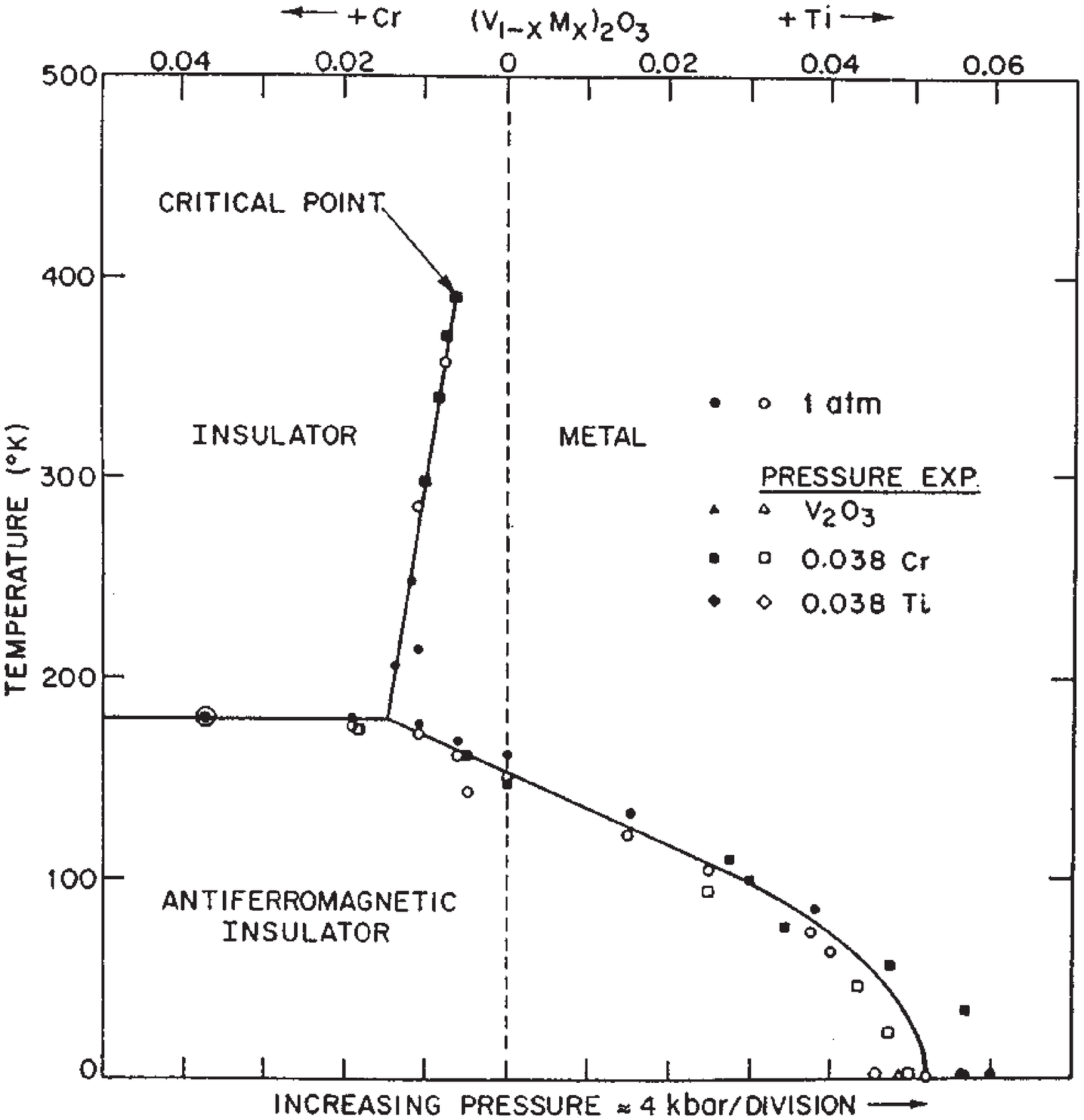}
%\vspace{7cm}

\caption{Experimental phase diagram of V$_2$O$_3$ doped with  Cr
and Ti [reproduced from Ref.\protect~\citen{McWahn}]. Doping
V$_2$O$_3$ effects the lattice constants in a similar way as
applying pressure (generated either by a hydrostatic pressure $P$,
or by changing the $V$-concentration from V$_2$O$_3$ to
V$_{2-y}$O$_3$) and leads to a Mott-Hubbard transition between the
{\em paramagnetic} insulator (PI) and metal (PM). At lower
temperatures, a Mott-Heisenberg transition between the
paramagnetic metal (PM) and the {\em antiferromagnetic} insulator
(AFI) is observed.\label{PDv2o3}}
\end{figure}

One of the most famous examples of a cooperative electronic
phenomenon occurring at intermediate coupling strengths is the
transition between a paramagnetic metal and a paramagnetic
insulator induced by the Coulomb interaction between the electrons
-- the Mott-Hubbard metal-insulator transition. The question
concerning the nature of this transition poses one of the
fundamental theoretical problems in condensed matter
physics.\cite{Mott} Correlation-induced metal-insulator
transitions (MIT) are found, for example, in transition metal
oxides with partially filled bands near the Fermi level. For such
systems band theory typically predicts metallic behavior. The most
famous example is V$_{2}$O$_{3}$ doped with Cr as shown in
Fig.~\ref{PDv2o3}. While at low temperatures V$_{2}$O$_{3}$
is an antiferromagnetic insulator with monoclinic crystal
symmetry, it has a corundum structure
 in the high-temperature paramagnetic phase. All
transitions shown in the phase diagram are of first order.  In the
case of the transitions from the high-temperature paramagnetic
phases into the low-temperature antiferromagnetic phase this is
naturally explained by the fact that the transition is accompanied
by a change in crystal symmetry. By contrast, the crystal symmetry
across the MIT in the paramagnetic phase remains intact, since
only the ratio of the $c/a$ axes changes discontinuously. This may
be taken as an indication for the predominantly electronic origin
of this transition which is not accompanied by any conventional
long-range order.
%as in the case of a Mott-Heisenberg transition,
%nor by a change of the crystal symmetry as in the case of a
%Peierls metal-insulator transition\cite{Mott}. Instead, it
From a models point of view the MIT is triggered by
%from electronic
%correlations or, more specifically,
a change of the ratio of the Coulomb interaction $U$ relative to
the bandwidth $W$.
%Experimentally this is achieved by doping
%V$_2$O$_3$ with, e.g., Cr or Ti, or by applying pressure $P$
%within the Cr doped phase.
Originally, Mott considered the extreme
limits $W=0$ (when atoms are  isolated and insulating) and $U=0$
where the system is metallic. While it is simple to describe these
limits, the crossover between them, i.e., the metal-insulator
transition itself, poses a very complicated electronic correlation
problem. Among others, this metal-insulator transition
 has been addressed by Hubbard in various approximations
\cite{Hubbard} and by Brinkman and Rice within the Gutzwiller
approximation \cite{BR}. During the last few years, our
understanding of the MIT in the one-band Hubbard model has
considerably improved, in particular due to the application of
dynamical mean-field theory \cite{DMFTMott}.

\begin{figure}[tb]
\centerline{
\includegraphics[clip=off,width=4.cm]{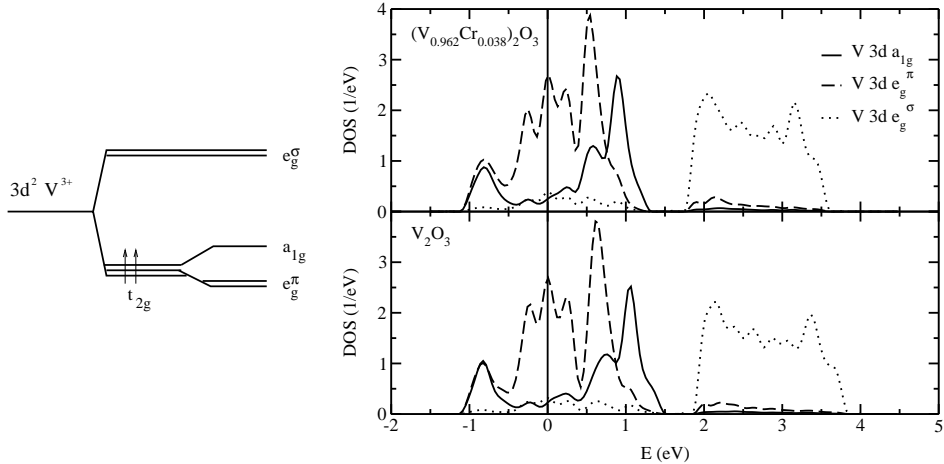} \hfill
\includegraphics[clip=off,width=8.cm]{gr_lda_big.eps} }
\caption{ Left: Scheme of the $3d$ levels in the corundum
crystal structure. Right: Partial LDA DOS of the 3d bands for paramagnetic
metallic $ {\rm V_2O_3}$ and insulating ${\rm
(V_{0.962}Cr_{0.038})_2O_3}$ [reproduced from Ref.\protect
\citen{Held01a}]. \label{V2O3ldados}} \vspace{-.9em}
\end{figure}

Both the paramagnetic {\em metal} V$_2$O$_3$ and the
paramagnetic {\em insulator} ${\rm %
(V_{0.962}Cr_{0.038})_{2}O_{3}}$ have the same corundum crystal
structure with only slightly different lattice parameters.
\cite{dernier70a,Foot1} Nevertheless, within LDA both phases are
found to be metallic (see Fig.~\ref{V2O3ldados}). The LDA DOS
shows a splitting of the five Vanadium d-orbitals into three $t_{2g}$ states
near the Fermi energy and two  $e_{g}^{\sigma }$ states at higher
energies. This reflects the (approximate) 
octahedral arrangement of oxygen around
    the vanadium atoms. Due to the trigonal symmetry of the
    corundum structure  the  $t_{2g}$ states
are further split into one $a_{1g}$ band and two degenerate
$e_{g}^{\pi }$ bands, see  Fig.~\ref{V2O3ldados}.
The only visible  difference between ${\rm %
(V_{0.962}Cr_{0.038})_{2}O_{3}}$ and ${\rm V_{2}O_{3}}$ is a
slight narrowing of the $t_{2g}$ and $e_{g}^{\sigma }$ bands by
$\approx 0.2$ and $0.1$ eV, respectively as well as a weak
downshift of the centers of gravity of both groups of bands for
${\rm V_{2}O_{3}}$. In particular, the insulating gap of the
Cr-doped system is seen to be missing in the LDA DOS. Here we will
employ LDA+DMFT(QMC) to show explicitly that the insulating gap is
caused by the electronic correlations.
In particular, we make use of the simplification for transition
metal oxides described in Section~\ref{SimpTMO} and restrict the
LDA+DMFT(QMC) calculation to the three $t_{2g}$ bands at the Fermi
energy, separated from the  $e_{g}^{\sigma }$ and oxygen bands.

\begin{figure}[tb]
\vspace{-10.1cm}

\begin{center}
\includegraphics[clip=off,width=9.cm]{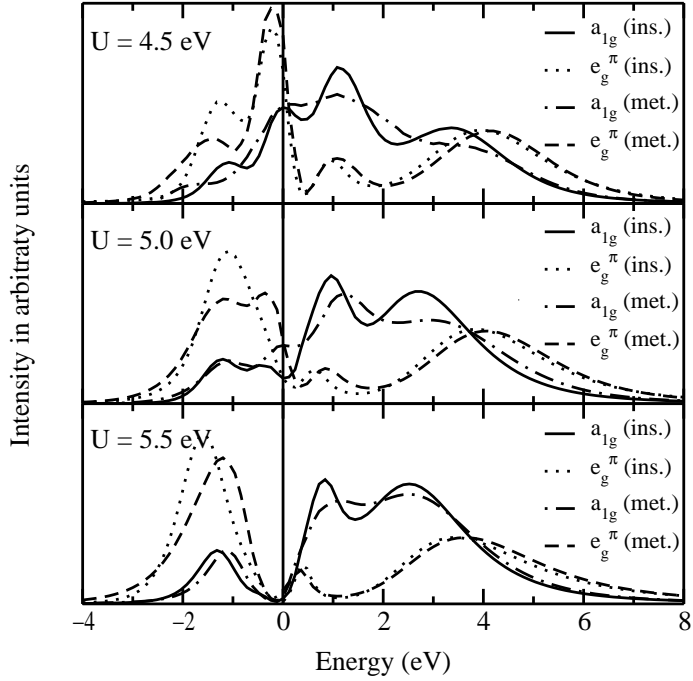}
\vspace{10.1cm}

\end{center}
\vspace{-.1em} \caption{LDA+DMFT(QMC) spectra for paramagnetic
${\rm (V_{0.962}Cr_{0.038})_2O_3}$ (``ins.'') and ${\rm V_2O_3} $
(``met.'') at $U=4.5$, $5$ and $5.5$ eV, and $T=0.1$ eV $\approx
1000$ K [reproduced from Ref.\protect \citen{Held01a}].}
\label{spectrum} \vspace{-.1em}\end{figure}

 While the Hund's rule coupling $J$ is
insensitive to screening effects and
may, thus, be obtained within LDA to a good accuracy ($J=0.93$ eV \cite%
{Solovyev}), the LDA-calculated value of the Coulomb repulsion $U$
has a typical uncertainty of at least 0.5 eV\cite{Nekrasov00}. To
overcome this uncertainty, we study the spectra obtained by
LDA+DMFT(QMC) for three different values of the Hubbard
interaction ($U=4.5, 5.0, 5.5$) in
Fig.~\ref{spectrum}.  
All QMC results presented were obtained for $T\!=\!0.1$ eV. However,
simulations for V$_2$O$_3$ at $U\!=\!5$ eV, $T\!=\!0.143$ eV, and 
$T\!=\!0.067$ eV 
suggest only a minor smoothing of the spectrum with increasing temperature.
From the results obtained we
conclude that the critical value of $U$ for the MIT is at about
$5$ eV: At $U=4.5$ eV one observes pronounced quasiparticle peaks
at the Fermi energy, i.e., characteristic metallic behavior, even
for the crystal structure of
the insulator ${\rm %
(V_{0.962}Cr_{0.038})_{2}O_{3}}$, while at $U=5.5$ eV the form of
the calculated spectral function is typical for an insulator for
both sets of crystal structure parameters. At $U=5.0$ eV one is
then at, or very close to, the MIT since there is a pronounced dip
in the DOS at the Fermi energy for both $a_{1g}$ and $e_{g}^{\pi
}$ orbitals for the crystal structure
of $%
{\rm (V_{0.962}Cr_{0.038})_{2}O_{3}}$, while for pure ${\rm
V_{2}O_{3}}$ one still finds quasiparticle peaks. (We note that at
$T\approx 0.1$ eV one only observes metallic-like and
insulator-like behavior, with a rapid but smooth\ crossover
between these two phases, since a sharp MIT occurs only at lower
temperatures\cite{Rozenberg97a,DMFTMott}). The critical value of
the Coulomb interaction $U\approx 5$ eV is in reasonable agreement
with the values determined spectroscopically by fitting to model
calculations, and by constrained LDA, see \cite{Held01a} for
details.

\begin{figure}[tb]
\centerline{
\includegraphics[width=4.8cm,height=5.5cm]{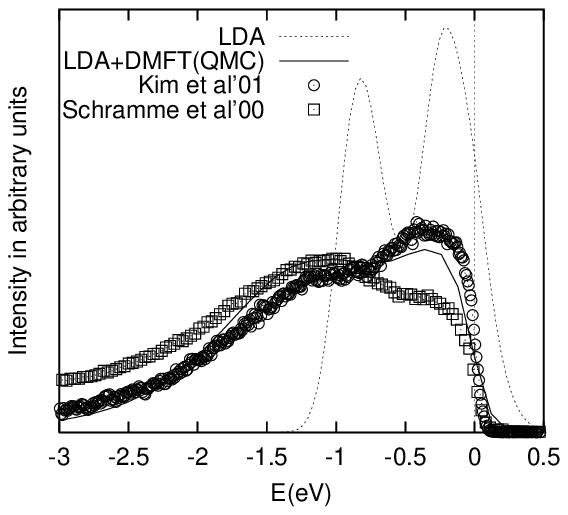}\hspace{-.4cm}
\includegraphics[width=8.8cm,height=5.5cm]{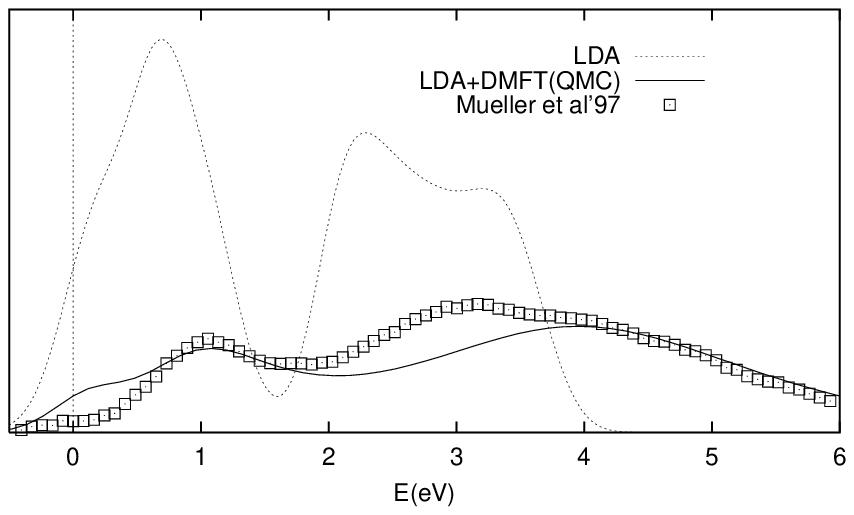} }
\caption{Comparison of the LDA+DMFT(QMC)
spectrum\protect\cite{Held01a} at  $U=5\,$eV and $T=0.1$ eV
$\approx 1000$ K below (left Figure) and above (right Figure) the
Fermi energy (at 0$\,$eV) with the LDA
spectrum\protect\cite{Held01a} and the
 experimental spectrum
(left: photoemission spectrum of  Schramme {\it et al.} \protect
\cite{Schramme00} and Kim {\it et al.} \protect\cite{Kim01};
right: X-ray absorption spectrum of  M\"{u}ller {\it et al.}
\protect\cite{Mueller}).} \label{compexp}
\end{figure}

To compare with the photoemission spectrum of ${\rm V_{2}O_{3}}$
spectrum by Schramme {\em et al.}\cite{Schramme00} and by Kim {\it
et al.}\cite{Kim01} as well as with the X-ray absorption data by
M\"{u}ller {\it et al.}\cite{Mueller}, the LDA+DMFT(QMC) spectrum
of Fig.~\ref{spectrum} is multiplied with the Fermi function  at
$T=0.1$ eV and Gauss-broadened by $0.05$ eV to account for the
experimental resolution. The theoretical result for $U=5$ eV is
seen to be in good agreement with experiment (Fig.~\ref{compexp}).
In contrast to the LDA results, our results not only describe the
different bandwidths above {\em and} below the Fermi energy
($\approx 6$ eV and $\approx 2-3$ eV, respectively), but also
 the position of two (hardly distinguishable) peaks below the Fermi
energy (at about -1$\,$eV and -0.3$\,$eV) as well as  the
pronounced  two-peak structure above the Fermi energy (at about
1$\,$eV and 3-4$\,$eV). While LDA also gives  two peaks below and
above the Fermi energy, their position and physical origin is
quite different. Within LDA+DMFT(QMC) the peaks at  -1$\,$eV
and 3-4$\,$eV are the incoherent Hubbard bands induced by the
electronic correlations whereas in the LDA the peak at 2-3$\,$eV
is caused by the $e_{g}^{\sigma}$ states and that at  -1$\,$eV is
the band edge maximum of the  $a_{1g}$ and $e_{g}^{\pi }$ states
(see Fig.\ \ref{V2O3ldados}). Note that the theoretical and
experimental spectrum is highly {\em asymmetric} w.r.t the Fermi
energy. This high {\em asymmetry} which is caused by the orbital
degrees of freedom is missing in the one-band Hubbard model which
was used  by Rozenberg {\it et al.}\cite{Rozenberg95} to
describe  the optical spectrum of ${\rm V_{2}O_{3}}$.

The comparison between theory and experiment for  Cr-doped {\em
insulating} ${\rm V_{2}O_{3}}$ is not as good as for
metallic ${\rm V_{2}O_{3}}$, see Ref.~\citen{Kim01}. 
This might  be, among other reasons,
due to the different Cr-doping of experiment and theory, the
difference in  temperatures (which is important because the
insulating gap of a Mott insulator is filled when increasing the
temperature \cite{DMFTMott}), or the fact that every V ion has a
unique neighbor in one direction, i.e., the LDA supercell
calculation has {\it a pair} of V ions per unit cell. The latter
aspect has so far not been included but arises naturally when one
goes from the simplified calculation scheme described in Section
\ref{SimpTMO} (and employed in the present Section with different
self-energies for the $a_{1g}$ and $e_{g}^{\pi }$ bands) 
to a full Hamiltonian calculation.

Particularly interesting are the spin and the orbital degrees of
freedom in ${\rm V_{2}O_{3}}$. From our
calculations\cite{Held01a}, we conclude that the spin state of
${\rm V_{2}O_{3}}$ is $S=1$ throughout the Mott-Hubbard transition
region. This agrees with the measurements of Park {\em et
al.}\cite{park} and also with the data for the high-temperature
susceptibility \cite{S-exp}.  But, it is at odds with the
$S\!=\!1/2$  model by Castellani {\it et al.}\cite{Castellani} and
with the results for a  one-band Hubbard model  which  corresponds
to $S\!=\!1/2$ in the insulating phase and, contrary to our
results, shows a  substantial change of the  local magnetic moment
at the MIT\cite{DMFTMott}. For the orbital degrees of freedom we find a
predominant occupation of
the $%
e_{g}^{\pi }$ orbitals, but with a significant admixture of
$a_{1g}$ orbitals. This admixture decreases at the MIT: in the
metallic phase we determine the occupation of the ($a_{1g}$,
$e_{g1}^{\pi }$, $e_{g2}^{\pi }$) orbitals as (0.37, 0.815,
0.815), and in the insulating phase as (0.28, 0.86, 0.86). This
should be compared with the experimental results of Park {\it et
al.}\cite{park}. From their analysis of the linear dichroism data
the authors concluded that the ratio of the configurations
$e_{g}^{\pi }e_{g}^{\pi }$:$e_{g}^{\pi }a_{1g}$ is equal to 1:1
for the paramagnetic metallic and 3:2 for the paramagnetic
insulating phase, corresponding to a one-electron occupation of
(0.5,0.75,0.75) and (0.4,0.8,0.8), respectively. Although our
results show a somewhat smaller value for the admixture of
$%
a_{1g}$ orbitals, the overall behavior, including the tendency of
a {\em
decrease }of the $a_{1g}$ admixture across the transition to the
insulating state, are well reproduced.

In the study above, the experimental crystal parameters of
  ${\rm V_{2}O_{3}}$ and  ${\rm (V_{0.962}Cr_{0.038})_{2}O_{3}}$
have been taken from the experiment. This leaves the question
unanswered whether a change of the lattice is the driving force
behind the Mott transition, or whether it is the electronic Mott
transition which  causes a change of
the lattice.
For another system, Ce, we will show
in  Section \ref{Ce} that the energetic changes near a Mott
transition are indeed sufficient to cause a first-order volume
change.

\section{The Cerium volume collapse: An example for a $4f$-electron system}
\label{Ce}

Cerium exhibits a transition from the  $\gamma$- to the $\alpha$-phase
with increasing pressure or decreasing
temperature. This transition is accompanied
by an unusually large volume change of 15\% \cite{EXPT},
much larger than the 1-2\%   volume change
in  ${\rm V_{2}O_{3}}$.  The $\gamma$-phase may also be prepared
in metastable form at room temperature in which case the reverse
$\gamma$-$\alpha$ transition occurs under pressure \cite{Olsen}.
Similar volume collapse transitions are observed under pressure in Pr
and Gd (for a recent review see Ref.~\citen{JCAMD}). 
 It is widely believed
that these transitions arise from changes in the degree of $4f$
electron correlation, as is reflected in both 
the Mott transition\cite{JOHANSSON} and the Kondo volume
collapse (KVC)\cite{KVC}  models.

The Mott transition model envisions a change from
itinerant, bonding character of the $4f$-electrons in the
$\alpha$-phase to non-bonding, localized character in the
$\gamma$-phase, driven by changes in the $4f$-$4f$ inter-site
hybridization.  
Thus, as the ratio of the $4f$ Coulomb interaction to
the $4f$-bandwidth increases, a Mott transition
occurs to the $\gamma$-phase, similar to the Mott-Hubbard transition of 
the 3d-electrons in ${\rm V_{2}O_{3}}$ (Section \ref{v2o3}).

The Kondo volume collapse\cite{KVC} scenario ascribes the collapse to a strong
change in the energy scale associated with the screening of the local
$4f$-moment by conduction electrons (Kondo screening), which is
accompanied by the appearance of an Abrikosov-Suhl-like quasiparticle
peak at the Fermi level. In this model the $4f$-electron spectrum of Ce
would change across the transition in a fashion very similar to the 
Mott scenario, i.e., a strong reduction of the spectral weight at the Fermi
energy should be observed in going from the $\alpha$- to the 
$\gamma$-phase. The subtle difference comes about by the $\gamma$-phase having
metallic $f$-spectra with a strongly enhanced effective mass
as in a heavy fermion system, in contrast to the $f$-spectra
characteristic of an insulator in the case of the Mott scenario.
The $f$-spectra in the Kondo picture also exhibit Hubbard side-bands
not only in the $\gamma$-phase, but in the $\alpha$-phase as
well, at least close to the transition.  While local-density and
static mean-field theories correctly yield the Fermi-level peaks
in the $f$-spectra for the $\alpha$-phase, they do not exhibit
such additional Hubbard side-bands, which is sometimes taken
as characteristic of the ``$\alpha$-like'' phase in the Mott
scenario \cite{JOHANSSON}.  However, this behavior is more likely
a consequence of the static mean-field treatment, as correlated
solutions of both Hubbard and periodic Anderson models exhibit
such residual Hubbard side-bands in the $\alpha$-like regimes
\cite{Held00a}.

Typically, the Hubbard model and the periodic Anderson model
are considered as paradigms for the Mott and KVC model, respectively.
Although both models describe completely different physical situations
it was shown recently that one can observe a surprisingly similar behavior at
finite temperatures: the evolution of the spectrum and the local 
magnetic moment with increasing Coulomb interaction
 show very similar features as well as, in the case
of a periodic Anderson model with nearest neighbor hybridization,
the phase diagram and the charge compressibility
\cite{Held00a,Held00b}. From this point of view the distinction between
the two scenarios appears to be somewhat  artificial, at least at temperatures
relevant for the description of the $\alpha$-$\gamma$ transition.

For a realistic  calculation of the Cerium $\alpha$-$\gamma$ transition,
we employ  the full Hamiltonian calculation described in
Sections \ref{lda}, \ref{ldaCoulomb}, and \ref{dmft} 
where the one-particle Hamiltonian was calculated by LDA
and the $4f$ Coulomb interaction $U$ along with the associated $4f$ site energy shift
by a constrained LDA calculation (for details 
of the the two independent calculations presented in the current Section
see Refs.~\citen{JCAMD,McMahan01} and Ref.~\citen{Zoelfl01}).
We have not included the spin-orbit interaction
which has a rather small impact on LDA results for Ce, nor the
intra-atomic exchange interaction which is less relevant for Ce as
occupations with more than one $4f$-electron on the same site are
rare. Furthermore, the $6s$-, $6p$-, and $5d$-orbitals are assumed to be non-interacting
in the formalism of Eq. (\ref{Hint}), Section \ref{ldaCoulomb}.
Note, that the $4f$ orbitals are even better localized than the $3d$ orbitals
and, thus, 
uncertainties in  $U$ 
are relatively small  and would only translate into a
possible volume shift for the $\alpha$-$\gamma$-transition.

\begin{figure}[tb]
\begin{center}
\hspace{-1.cm} \epsfxsize=9cm \epsfbox{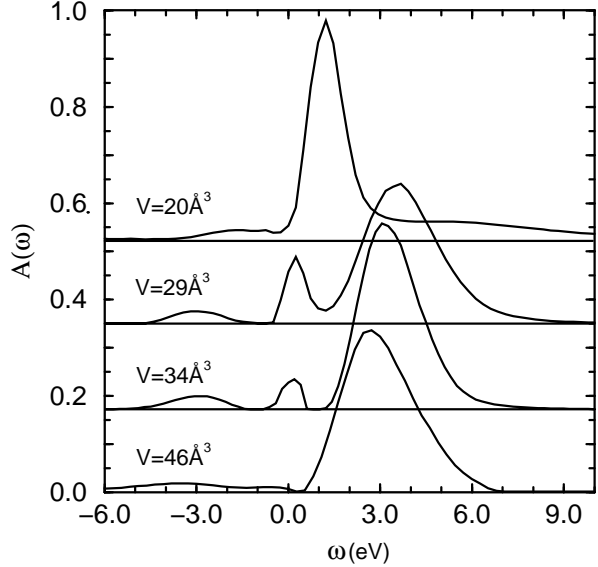}
\end{center}
\caption{
Evolution of the $4f$ spectral function $A(\omega)$ with
volume at $T\!=\!0.136\:$eV ($\omega\!=\!0$ corresponds to the chemical potential;
curves are offset as indicated; $\Delta
\tau\!=\!0.11\!$$\,$eV$^{-1}$).
Coinciding with the sharp anomaly in the correlation
energy (Fig.~\ref{figE}), the central quasiparticle resonance   disappears,
at least at finite temperatures [reproduced
from Ref.~\citen{McMahan01}].
\label{figSpecCe}}
\end{figure}

The LDA+DMFT(QMC) 
spectral evolution of the Ce 
$4f$-electrons is  presented in Fig.~\ref{figSpecCe}.
It shows similarities to ${\rm V_{2}O_{3}}$  (Fig.~\ref{spectrum}, Section \ref{v2o3}):
 At a volume per atom $V\!=\!20\,$\AA$^3$, Fig.~\ref{figSpecCe} shows that almost the entire
spectral weight lies in a large quasiparticle peak with a center of
gravity slightly above the chemical potential.  This is similar to
the LDA solution, however, a weak upper Hubbard band is also
present even at this small volume.  At the volumes $29$$\,$\AA$^3$ and
$34$$\,$\AA$^3$ which approximately bracket the $\alpha$-$\gamma$
transition, the spectrum has a three peak structure.
 Finally, by $V\!=\!46$$\,$$\,$\AA$^3$, the central peak has
disappeared leaving only the lower and upper Hubbard bands.
However, an important  difference to ${\rm V_{2}O_{3}}$ is that 
the $spd$-spectrum shows metallic behavior 
and, thus, Cerium  remains a metal throughout
this transition monitored by a vanishing $4f$ quasiparticle resonance.

To study the energetic changes associated with
the rapid change of the quasiparticle weight at the Fermi energy, we
calculate the DMFT energy per site for the model Hamiltonian  (\ref{Hint})
\begin{eqnarray}
E_{\rm DMFT}&\! =\!& \frac{T}{N_k} \sum_{n {\bf k} \sigma}
{\rm Tr}({ H}^0_{\rm LDA}({\bf k}) { G}_{\bf k}(i\omega_n))
e^{i\omega_n 0^+} + U_f \, d.
\label{Eng}
\end{eqnarray}
Here, Tr denotes the trace over the $16\times16$ matrices, $T$ the
temperature, $N_k$ the number of ${\bf k}$ points, $G_{\bf k}$ the Green function matrix w.r.t.
the orbital indices, ${ H}^0_{\rm LDA}({\bf k})$ the LDA one-particle matrix
Eq.\ (\ref{LDAHam}),
and 
\begin{equation}
d = \frac12
{\sum}_{m\sigma,m'\sigma'}' \langle \hat{n}_{ifm\sigma}\,
\hat{n}_{ifm'\sigma'}\rangle
\label{double}
\end{equation}
is a generalization of the one-band
double occupation for multi-band models.

\begin{figure}[tb]
\begin{center}
 \epsfxsize=10cm \epsfbox{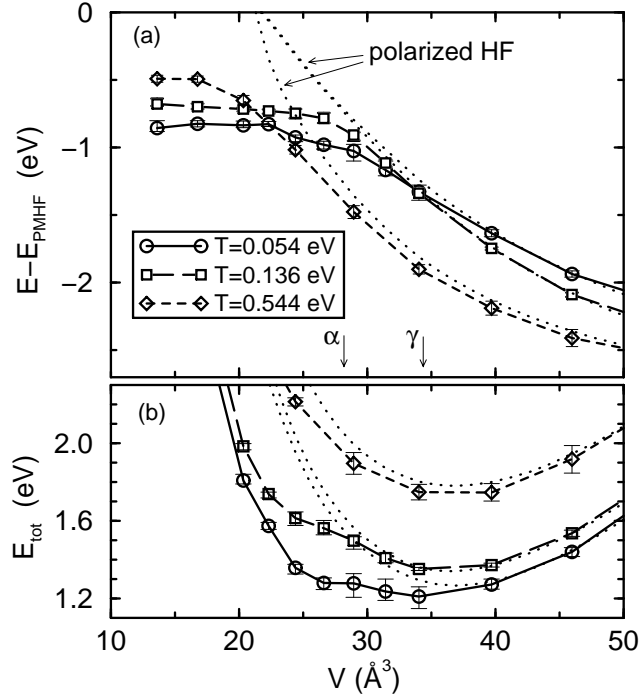}
\end{center}

\caption{(a) Correlation energy $E_{\rm DMFT}\!-\!E_{\rm PMHF}$
as a function of atomic volume (symbols) and 
polarized HF
energy $E_{\rm AFHF}\!-\!E_{\rm PMHF}$ (dotted lines which, at large V, approach the
DMFT curves for the respective temperatures);
arrows:  observed volume collapse from the $\alpha$- to the $\gamma$-phase.  
The correlation energy sharply bends away from the
polarized HF energy in the region of the transition. (b) The resultant negative curvature leads to a growing depression
     of the total energy near $V\!=\!26$--$28$ \AA$^3$ as temperature
     is decreased, consistent with an emerging double well at still
     lower temperatures and thus the $\alpha$-$\gamma$ transition. The
     curves at $T=0.544\,$eV were shifted downwards in (b)  by $-0.5\,$eV to match the energy range [reproduced
from Ref.~\citen{McMahan01}].
\label{figE}}
\end{figure}

Fig.~\ref{figE}a shows our calculated DMFT(QMC) energies $E_{\rm DMFT}$
as a function of atomic volume at three temperatures  {\it relative}
to the paramagnetic Hartree Fock (HF) energies $E_{\rm PMHF}$ [of the Hamiltonian (\ref{Hint})], 
i.e., the energy contribution
due to {\em electronic correlations}. 
Similarly given are the polarized HF energies
which reproduce $E_{\rm DMFT}$
at large volumes and low temperatures.
With decreasing  volume, however,  the DMFT
energies bend away from the polarized HF solutions.
Thus,  at $T\!=\!0.054\,$eV$\,\approx 600\,$K, 
a region of negative curvature in $E_{\rm
DMFT}\!-\!E_{\rm PMHF}$ is evident within the observed two phase region (arrows).

Fig.~\ref{figE}b presents the calculated LDA+DMFT total energy $E_{\rm
tot}(T)\!=\!E_{\rm LDA}(T)\!+\!E_{\rm DMFT}(T)\!-\!E_{\rm mLDA}(T)$
where $E_{\rm mLDA}$ is the energy of an LDA-like solution of the
Hamiltonian (\ref{Hint}) \cite{mLDA}.  Since both $E_{\rm LDA}$
and $E_{\rm PMHF}\!-\!E_{\rm mLDA}$ have positive curvature throughout
the volume range considered, it is the negative curvature of the
correlation energy in Fig.~\ref{figE}a which leads to the dramatic
depression of the LDA+DMFT total energies in the range 
$V\!=\,$26-28$\,$\AA$^3$ for decreasing temperature, which 
contrasts to the smaller
changes near $V\!\!=\!34\,$\AA$^3$ in Fig.~\ref{figE}b.  This trend is
consistent with a double well structure emerging at still lower
temperatures (prohibitively expensive for QMC simulations), and with it
a first-order volume collapse.
 This is in
reasonable agreement with the experimental volume collapse
 given our use of energies rather
than free energies, the different temperatures, and the
LDA and DMFT approximations.
A similar scenario has been proposed recently for the
$\delta$-$\alpha$ transition in Pu on the basis of
LDA+DMFT calculations \cite{SAVRASOV}, which solves DMFT
by an ansatz inspired by  IPT  and includes a modification of the
DFT/LDA step to account for the density changes introduced by the DMFT\cite{SAVRASOV2}.

In a separate LDA+DMFT(NCA) calculation for Ce, we have obtained
a number of physical quantities for both phases which may be
compared to experimental values \cite{Zoelfl01}. Various static properties extracted
\begin{table}[tb]
%\squeezetable
\begin{center}
\begin{tabular}{|p{3.2cm}|p{1.5cm}|p{1.75cm}|p{1.5cm}|p{1.5cm}|}
\hline   & $\alpha$-Ce$^{\rm Theo}$& $\alpha$-Ce\cite{Liu,Murani}  
         & $\gamma$-Ce$^{\rm Theo}$& $\gamma$-Ce \cite{Liu,Murani}\\
\hline \centering $P_0$ & 0.126 & 0.1558&  0.0150 & 0.0426\\ 
\hline \centering $P_1$ & 0.829 & 0.8079&  0.9426 & 0.9444\\ 
\hline \centering $P_2$ & 0.044 & 0.0264&  0.0423 & 0.0131\\
\hline \centering $n_f$ & 0.908 & $0.8\ldots0.861$& 1.014 & $0.971\ldots1$\\
\hline \centering $T_K$,~\eka{K}\ekz & 1000 & $945\ldots2000$ & 30 & $60\ldots95$\\
\hline \centering $\chi$,~\eka{$10^{-3}emu/mol$}\ekz & 1.08  & $0.53\ldots0.70$ & 24  & $8.0\ldots12$ \\
\hline 
\end{tabular}
\end{center}
\caption{Comparison between LDA+DMFT(NCA) calculated parameters for both 
$\alpha$-  and $\gamma$-phase at $T=580~K$ and experimental values\cite{Liu,Murani} [reproduced
from Ref.~\citen{Zoelfl01}]. $P_0$, $P_1$ and $P_2$ are partial probabilities
for an empty, singly and doubly occupied $4f$-state, $n_f$ is the $f$-electron
occupancy,
$T_{\rm K}$ the estimated Kondo temperature, and $\chi$ the magnetic
susceptibility.\label{tab:cer}
}
\end{table}
from the calculations\cite{Zoelfl01} and their counterparts from experiments are collected in
Table~\ref{tab:cer} and show an overall fair to good agreement in the tendencies
and, except for the susceptibility, the
absolute values. Since the calculation of
the magnetic susceptibility
$\chi$ in Ref.~\citen{Zoelfl01} was based on simplifying assumptions, the absolute
numbers cannot be expected to match experiment. However, the general tendency
and especially the ratio between $\alpha$- and $\gamma$-Ce is in good agreement
with experiment. It is interesting to note that the experiments predict a
\begin{figure}[tb]
\begin{center}\mbox{}
 \epsfxsize=9cm \epsfbox{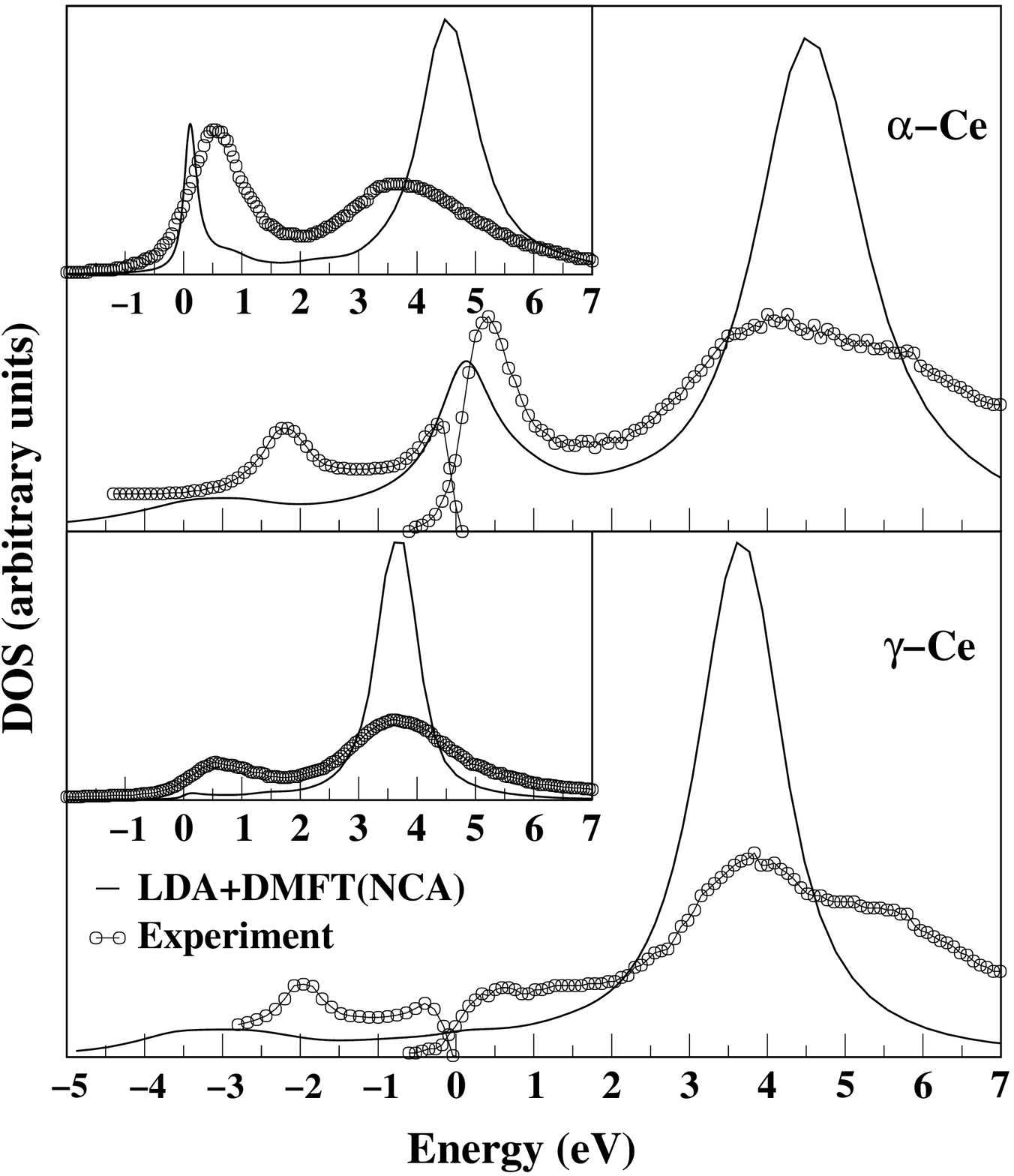}
\end{center}
\caption{Comparison between combined photoemission~\protect\cite{PES_Wieliczka}
and BIS~\protect\cite{BIS_Wuilloud} experimental (circles) and theoretical 
(solid line) $f$-spectra for $\alpha$- (upper part) and $\gamma$-Ce 
(lower part) at $T=580\,$K. The relative intensities of the BIS and photoemission
portions are roughly for one 4$f$ electron. The experimental and 
theoretical spectra were normalized and the theoretical curve was 
broadened with resolution width of $0.4\,$eV.
In the insets a comparison between RIPES~\protect\cite{Grioni}
experimental (circles) and theoretical (solid line) $f$-spectra is given.
The experimental and theoretical data were normalized and the theoretical 
curve was broadened with broadening coefficient of $0.1\,$eV [reproduced
from Ref.~\citen{Zoelfl01}].
\label{fig:CeSpec}}
\end{figure}
finite Kondo screening-scale for both phases, which actually would point
toward the KVC scenario. Finally, let us compare spectral functions for the
$4f$-states calculated with the LDA+DMFT(NCA) approach to experimental data\cite{PES_Wieliczka}.
The photoemission  spectrum for $\alpha$-Ce (upper part of Fig.~\ref{fig:CeSpec})
shows a main structure between $3\:$eV and $7\:$eV,
which is attributed to $4f^2$ final state multiplets. In the calculated 
spectrum all excitations to $4f^2$ states are described by the featureless
upper Hubbard band. As a consequence of the simplified interaction model 
all doubly occupied states are degenerate. This shortcoming in our 
calculation is responsible for the sharply peaked main structure. 
The neglected exchange interaction would produce a multiplet structure, which
would be closer to the experiment. 
The experimental peak at about $0.5\:$eV is attributed to two $4f^1$ final
states, which are split by spin-orbit coupling. 
The calculated $f$-spectrum shows a sharp quasiparticle or Kondo resonance 
slightly above the Fermi energy, which is the result 
of the formation of a singlet state between $f$- and 
conduction states.   
We thus suggest that the spectral weight seen in the experiment is a result
of this quasiparticle  resonance. Since we did not yet include spin-orbit coupling
in our model, we  cannot observe the mentioned splitting of the 
resonance. However, as it is well known~\cite{Kondo_degbreak}, the 
introduction of such a splitting would eventually split the 
Kondo resonance. If we used the experimentally determined value of about 
$0.3\:$eV for the spin-orbit splitting \cite{BIS_Wuilloud}, 
the observed resonance of width $0.5\:$eV would indeed occur in the 
calculations.
In the lower part of Fig.~\ref{fig:CeSpec}, 
a comparison between experiment and our calculation 
for $\gamma$-Ce is shown. The most striking difference between lower and upper 
part of Fig.~\ref{fig:CeSpec}
is the absence of the  Kondo resonance in the high temperature phase 
($\gamma$-Ce; transition temperature $141~K$~\cite{EXPT}) which is in agreement
with our calculations.

In the insets of Fig.~\ref{fig:CeSpec}, our results for the non-occupied states
in the $f$-density are compared with RIPES data~\cite{Grioni}.
The calculated $f$-spectra were
 multiplied by the Fermi-step function and broadened with an 
Lorentzian of the width $0.1\,$eV
 in order to mimic the experimental resolution 
in the theoretical curves. Here, as above the theoretical overestimation of 
the sharpness of the
upper Hubbard band is a consequence of the simplified local interaction
and thus of the missing multiplet structure of the $4f^2$-final states.
The main feature of the experimental spectra, i.e.,
a strong decrease
of the intensity ratio for Kondo resonance  and upper Hubbard band peaks 
from $\alpha$- to $\gamma$-Ce, can also be seen in 
the theoretical curves
of Fig.~\ref{fig:CeSpec} as well as in the
study presented in  Fig.~\ref{figSpecCe}. A more thorough comparison
of these two independent LDA+DMFT(NCA) and  LDA+DMFT(QMC) studies
remains to be done.

\section{Conclusion and Outlook}
\label{conclusion}
In this paper we discussed the set-up of the computational scheme
LDA+DMFT
which merges two non-perturbative, complementary investigation
techniques
for many-particle systems in solid state physics. LDA+DMFT allows one to
perform \emph{ab initio} calculations of real materials with strongly
correlated electrons. Using the band structure results calculated within
local density approximation (LDA) as input, the missing electronic
correlations are introduced by dynamical mean-field theory (DMFT). On a
technical level this requires the solution of an effective
self-consistent,
multi-band Anderson impurity problem by some numerical method (e.g. IPT,
NCA, QMC). Comparison of the photoemission spectrum of
La$_{1-x}$Sr$_{x}$TiO$%
_{3}$ calculated by LDA+DMFT using IPT, NCA, and QMC reveal that the
choice
of the evaluation method is of considerable importance. Indeed, only
with
the numerically exact QMC quantitatively reliable results are obtained.
The
results of the LDA+DMFT(QMC) approach were found to be in very good
agreement with the experimental photoemission spectrum of
La$_{0.94}$Sr$%
_{0.06}$TiO$_{3}$.

We also presented results of a LDA+DMFT(QMC) study\cite{Held01a} of the
Mott-Hubbard metal-insulator transition (MIT) in the paramagnetic phase
of
(doped) $\mathrm{V_{2}O_{3}}$. These results showed a Mott-Hubbard MIT
at a
reasonable value of the Coulomb interaction $U\approx 5$eV and are in
very
good agreement with the experimentally determined photoemission and
X-ray
absorption spectra for this system, i.e., above \emph{and} below the
Fermi
energy. In particular, we find  a spin state $S=1$ in the
paramagnetic phase, and an orbital admixture of $e_{g}^{\pi }e_{g}^{\pi
}$
and $e_{g}^{\pi }a_{1g}$ configurations, which both agree with recent
experiments. Thus, LDA+DMFT(QMC) provides a remarkably accurate
microscopic
theory of the strongly correlated electrons in the paramagnetic metallic
phase of $\mathrm{V_{2}O_{3}}$.

%% BEGIN LaMnO3
%Transition metal oxides are not only characterized by physical features
%which are typical for the vicinity of a Mott-Hubbard MIT, but also
%display a
%rich spectrum of ordered phases. It is desirable for a theoretical
%technique
%to be able to describe these ordered phases, too. As an important
%example we
%presented LDA+DMFT(NCA) studies of the orbital and magnetic order in
%LaMnO$%
%_{3}$, using a simplified description of this system by neglecting the
%influence of the $t_{2g}$ states. We found the correct types of orbital
%and
%magnetic order with reasonable transition temperatures. Furthermore, we
%were
%able to show that an important aspect of both orbital and magnetic order
%is
%a strong redistribution of spectral weight in the local density of
%states
%(DOS). A comparison of the calculated orbital order parameter [WITH\
%WHAT?]
%showed a clear enhancement when magnetic order sets in, in accordance
%with
%recent resonant $x$-ray scattering experiments.\cite{Mur98} %% END
%LaMnO3

Another material where electronic correlations are considered to be
important is Cerium. We reviewed our recent investigations of the Ce
$\alpha
$-$\gamma $ transition, based on LDA+DMFT(QMC)\cite{McMahan01} and
LDA+DMFT(NCA)\cite{Zoelfl01} calculations. The spectral results and
susceptibilities show the same tendency as seen in the experiment,
namely a
dramatic reduction in the size of the quasiparticle peak at the Fermi
level
when passing from the $\alpha$- to the $\gamma$-phase. 
%This evolution
%of
%the $4f$-electron spectral function in Ce with respect to a change of
%the
%lattice volume is similar to that of the $3d$-electrons in $\mathrm{%
%V_{2}O_{3}}$, at least at the experimentally relevant temperatures.
%Other
%aspects are different [FROM\ WHAT- $\mathrm{V_{2}O_{3}?}$] as Ce remains
%metallic due to the $spd$-electrons and the fact that both $\alpha $ and
%as
%well as $\gamma $ Ce have a quasiparticle resonance at the Fermi energy.
While we do not know at the moment whether the  zero-temperature
quasiparticle peak will
completely disappear at an even larger volume 
(i.e., in a rather Mott-like
fashion) or simply fade away continuously with increasing volume 
(i.e., in a more Kondo-like fashion), an important aspect of 
our results is that the
rapid reduction in the size of the peak seems to coincide with the
appearance of a negative curvature in the correlation energy and a
shallow minimum in the total energy. 
%This suggest an
%emerging
%double-well in the total energy, i.e., a first-order volume collapse
%between
%the two minima in agreement with the experimental $\alpha $- and $\gamma
%$-
%phase volumes. In other words, 
This suggest that the electronic correlations responsible
for the reduction of the quasiparticle peak are associated with energetic
changes that are strong enough to cause a volume collapse in the sense
of
the Kondo volume collapse model\cite{KVC}, or a Mott transition
model\cite%
{JOHANSSON} including electronic correlations.

%% BEGIN XRuO4
%Furthermore we investigated the evolution of the multi-band electronic
%structure for the isoelectronic alloy series Ca$_{2-x}$Sr$_{2}$RuO$_{4}$
%at
%the intermediate doping ${2>x>0.5}$. Starting from the good metal
%Sr$_{2}$RuO%
%$_{4}$, we find the effect of Ca-substitution to be a transfer of
%electrons
%from the wider $xy$-band to the narrower $(xz,yz)$ bands until, at a
%critical value of $x_{c}=0.5,$ integer occupancy of both subbands is
%reached. The progressive rotation of the RuO$_{6}$ octahedra in this
%region
%leads to Mott localization of the three electrons in the narrower
%$(xz,yz)$
%bands while the wider $xy$-band which is then half-filled, remains
%metallic.
%This partial localization of the 4d electrons can explain the puzzling
%observation of the coexistence of free $S=1/2$ local moments and
%metallic
%behavior in Ca$_{1.5}$Sr$_{0.5}$RuO$_{4}$. In this region strong
%correlations, but no long-range order, are presented so that both LDA
%and
%LDA+U are inapplicable. At this concentration we employed the
%LDA+DMFT(NCA)
%approach. These calculational scheme give us reliable information on the
%evolution of the electronic distribution among the three orbitals in the
%$%
%t_{2g}$-subshell, which is the key to an understanding of the electronic
%properties of this system\cite{Anisimov01}. %% END XRuO4

At present LDA+DMFT is the only available \emph{ab initio}
computational technique which is able to treat correlated
electronic systems close to a Mott-Hubbard MIT, heavy fermions,
and $f$-electron materials. The physical properties of such
systems are characterized by the correlation-induced generation of
small, Kondo-like energy scales which require the application of
genuine many-body techniques. The appearance of Kondo-like energy
scales in strongly correlated systems leads to several
experimentally relevant consequences. One of the most important
features is the enhancement of the quasiparticle mass m$^{\ast }$
(i.e., the decrease of the quasiparticle residue Z). This
phenomenon can be observed as an enhancement of the coefficient
$\gamma$ in the specific heat. Another important characteristic
is the Wilson ratio between $\gamma$ and the Pauli spin
susceptibility $\chi$. 
Future LDA+DMFT investigations will determine 
these quantities for real systems, as well as
the optical conductivity, phase-diagrams, 
the local vertex function, and various susceptibilities.

LDA+DMFT provides, at last, a powerful tool for \emph{ab initio}
investigations of real materials with strong electronic
correlations.
Indeed, LDA+DMFT depends on the input from both band structure theory
\emph{%
and} many-body approaches. Hence, for this computational scheme to be
entirely successful in the future two strong and vital communities will
finally have to join forces.

\section*{Acknowledgments}

We are grateful to  J.~W.~Allen, P.~W.~Anderson, R.~Bulla, 
R. Claessen, U. Eckern, G.~Esirgen, A.~Georges,
 K.-H. H\"{o}ck, S. Horn, M.~Jarrell, J. Keller,  H.-D. Kim, D.~E. Kondakov, 
G.~Kotliar, J. L\ae gsgaard, A. Lichtenstein,
D.~van der Marel, T.~M. Rice,
 G.~A.~Sawatzky, J. Schmalian, M.~Schramme, M.~Sigrist, M.\ Ulmke,
 and M. Z\"olfl for helpful discussions.
We thank  A.~Sandvik for making available his maximum
entropy code. The QMC code of Ref.~\citen{georges96}  App.~D
was modified for use for some of the results of Section \ref{Ce}.
This work was supported in part by the
Deutsche Forschungsgemeinschaft through Sonderforschungsbereich
484 (DV,GK,VE), 
Forschergruppe HO 955/2 (VE), and project Pr 298/5-1 \& 2 (TP),
the Russian Foundation for Basic Research by RFFI-01-02-17063 (VA,IN),
the U.S.~Department of Energy by University California LLNL under
 contract No.
W-7405-Eng-48. (AM), the U.S.~National Science Foundation by
DMR-9985978 (RS), a Feodor-Lynen grant of the Alexander von Humboldt foundation (KH),
the Lorentz Center in Leiden, 
the Leibniz-Rechenzentrum, M\"{u}nchen, and the John v.~Neumann-Institut
for Computing, J\"{u}lich.

%\vspace{-1.em}
%
%
%  References
%


\begin{thebibliography}{10}
\bibitem{Note}
At present, QMC simulations of the DMFT equations
 are not feasible at the experimental temperature
(80K). We note, however, that no
intrinsic temperature dependence was observed in the experiment\cite{fujimori},
at least up to room temperature.
\bibitem{Foot1} Use of the crystal structure of Cr-doped ${\rm V_{2}O_{3}}$
for the insulating phase of pure ${\rm V_{2}O_{3}}$ is justified
by the observation that Cr-doping is equivalent to the application
of (negative) pressure.
\bibitem{mLDA}
 We solve self-consistently for 
$n_f$ using a $4f$ self energy $\Sigma =U_f(n_f-\frac12)$, and then 
remove this contribution from the
eigenvalue sum to get the kinetic energy.  The potential energy is
taken to be $\frac12 U_f n_f(n_f-1)$.
\end{thebibliography}
\end{document}